\documentclass[useAMS,usenatbib]{mnras}

\usepackage[dvips]{graphics}
\usepackage{graphicx}
\usepackage{amsmath}
\usepackage{epsfig}
\usepackage[latin1]{inputenc}
\usepackage{hyperref}

\usepackage[T1]{fontenc}
\usepackage{ae,aecompl,url}
\usepackage{multicol,placeins}
\usepackage{cuted,mathtools}

\newcommand{\kms}{km\,s$^{-1}$ } 
\newcommand{\kmsns}{km\,s$^{-1}$} 
\newcommand{\kmsikpc}{km\,s$^{-1}$\,kpc$^{-1}$ } 
\newcommand{\kmsikpcns}{km\,s$^{-1}$\,kpc$^{-1}$} 
\newcommand{\kmskpc}{km\,s$^{-1}$\,kpc }
\newcommand{\kmskpcns}{km\,s$^{-1}$\,kpc}

\title[GALAH velocity distributions]{The GALAH Survey: Stellar streams and how stellar velocity
distributions vary with Galactic longitude, hemisphere and metallicity}

\author[Quillen et al.]
{
Alice C. Quillen$^{1,2}$, 
Gayandhi De Silva$^{3,4}$,
Sanjib Sharma$^{4}$,
Michael Hayden$^{4,5}$,
\newauthor
Ken Freeman$^6$,
Joss Bland-Hawthorn$^{4,5,7}$, 
Maru\v sa \v Zerjal$^6$,  
\newauthor
Martin Asplund$^{5,6}$,
Sven Buder$^8$, 
Valentina D'Orazi$^{9}$,
Ly Duong$^6$,
\newauthor
Janez Kos$^4$,
Jane Lin$^6$,
Karin Lind$^{8,10}$,
Sarah Martell$^{11}$, 
\newauthor
Katharine Schlesinger$^6$,
Jeffrey D. Simpson$^{3}$,
Daniel B. Zucker$^{3,12,13}$,
Tomaz Zwitter$^{14}$, 
\newauthor
Borja Anguiano$^{12,15}$,
Daniela Carollo$^{5,16}$,
Luca Casagrande$^{6}$,
Klemen Cotar$^{14}$, 
\newauthor
Peter L. Cottrell$^{17,18}$,
Michael Ireland$^{6}$,
Prajwal R. Kafle$^{19}$,  
Jonathan Horner$^{20}$,
\newauthor
Geraint F. Lewis$^4$,
David M. Nataf$^{21}$,
Yuan-Sen Ting$^{22,23,24}$,
Fred Watson$^{3}$,
\newauthor
Rob Wittenmyer$^{25,26}$, \&
Rosemary Wyse$^{21}$
\\
$^1$Department of Physics and Astronomy, University of Rochester, Rochester, NY 14627, USA\\
$^2$Simons Fellow in Theoretical Physics\\
$^3$Australian Astronomical Observatory, 105 Delhi Rd, North Ryde, NSW 2113, Australia\\
$^4$Sydney Institute for Astronomy, School of Physics, A28, The University of Sydney, Sydney NSW 2006, Australia\\
$^5$Australian Research Council Centre of Excellence for All Sky Astrophysics in 3 Dimensions (ASTRO-3D)\\
$^6$Research School of Astronomy and Astrophysics, Australian National University,  Cotter Road,  Canberra, ACT 72611, Australia\\
$^7$Miller Professor, Department of Astronomy, Campbell Hall, UC Berkeley, CA 94720, USA\\
$^8$Max-Planck-Institut for Astronomy, K\"onigstuhl 17, D-69117 Heidelberg, Germany\\
$^{9}$Istituto Nazionale di Astrofisica, Osservatorio Astronomico di Padova, vicolo dell'Osservatorio 5, 35122, Padova, Italy\\
$^{10}$Department of Physics and Astronomy, Uppsala University, Box 516, SE- 751 20 Uppsala, Sweden\\
$^{11}$School of Physics, University of New South Wales, Sydney, NSW 2052, Australia\\
$^{12}$Department of Physics and Astronomy, Macquarie University, Sydney, NSW 2109, Australia\\
$^{13}$Research Centre in Astronomy, Astrophysics \& Astrophotonics, 
   Macquarie University, Sydney, NSW 2109, Australia\\
$^{14}$Faculty of Mathematics and Physics, University of Ljubljana, Jadranska 19, 1000 Ljubljana, Slovenia\\
$^{15}$Department of Astronomy, University of Virginia, P.O. Box 400325, 
   Charlottesville, VA 22904-4325, USA\\
$^{16}$Istituto Nazionale di Astrofisica, 
 Via dell'Osservatorio 20, Pino Torinese 10025,  Italy\\
$^{17}$School of Physical and Chemical Sciences, University of Canterbury, Christchurch, New Zealand\\
$^{18}$Monash Centre for Astrophysics, School of Physics and Astronomy, Monash University,  
          Clayton 3800, Victoria, Australia\\
$^{19}$ICRAR, The University of Western Australia, 35 Stirling Highway, Crawley, WA 6009, Australia\\
$^{20}$Division of Research and Innovation, University of Southern Queensland,
                Toowoomba, Queensland 4350, Australia\\
$^{21}$Center for Astrophysical Sciences and Department of Physics and Astronomy, The Johns Hopkins University, Baltimore, MD 21218, USA\\
$^{22}$Institute for Advanced Study,  Princeton, NJ 08540, USA\\
$^{23}$Department of Astrophysical Sciences, Princeton University, Princeton, NJ 08544, USA\\
$^{24}$Observatories of the Carnegie Institution of Washington, 813 Santa Barbara Street, Pasadena, CA 91101, USA\\ 
$^{25}$University of Southern Queensland, Computational Engineering and Science Research Centre, Toowoomba, Queensland 4350, Australia\\
$^{26}$Australian Centre for Astrobiology, University of New South Wales, Sydney, NSW 2052, Australia\\
}

\hypersetup{draft}

\begin{document}
\maketitle

\begin{abstract}
Using GALAH survey data of nearby stars, we look at how structure in the planar ($u,v$) velocity distribution depends on metallicity and on viewing direction within the Galaxy.  In nearby stars with distance $d \la 1$ kpc,   the Hercules stream is most strongly seen  in higher metallicity stars [Fe/H]$ > 0.2$. The Hercules stream peak $v$ value depends on viewed galactic longitude, which we interpret as due to the gap between the stellar stream and more circular orbits being associated with a specific angular momentum value of about 1640 \kmskpcns.   The association of the gap with a particular angular momentum value supports a bar resonant model for the Hercules stream.

Moving groups previously identified in {\it Hipparcos} observations are easiest to see in stars nearer than 250 pc, and their visibility and peak velocities in the velocity distributions depends on both viewing direction (galactic longitude and hemisphere) and metallicity. We infer that there is fine structure in local velocity distributions that varies over distances of a few hundred pc in the Galaxy.   

\end{abstract}

\begin{keywords}
Galaxy: disc -- 
Galaxy: kinematics and dynamics --
Galaxy: evolution  
\end{keywords}

\section{Introduction}

Stars in the disc of the Milky Way  can be described in terms of 
their distributions in space, velocity, age, and chemical abundances. 
Star formation takes place in gas structures producing young compact clusters, 
moving groups and stellar associations
that evaporate, disrupt and are dispersed across the Galaxy,
while leaving imprints in velocity and chemical abundance distributions 
\citep{eggen58a,eggen58b,eggen58c,eggen65a,eggen65b,eggen65c,eggen70,eggen95,eggen96,eggen98,freeman02,baumgardt03,desilva07}.
Non-axisymmetric perturbations from spiral arms, a bar or
satellite galaxies can also induce gaps, over-densities or dynamical streams
\citep{kalnajs91,weinberg94,dehnen99b,dehnen00,fux01,quillen03,quillen05,chakrabarty08,quillen09,gardner10,minchev10,quillen11,lepine11,grand15}
that can be seen in {\it local} velocity distributions that are
constructed from stars restricted to small regions or neighbourhoods in the Galaxy.   

Because streams and clumps are seen
in the velocity distribution of stars in the Solar neighbourhood  
\citep{kapteyn1905,dehnen98,skuljan99,famaey05}, this distribution is poorly 
described by a single velocity ellipsoid \citep{schwarzschild1907,robin17,anguiano18}.
If clumps or gaps in the velocity distribution of old stars in the disc are due solely to resonances
with spiral or bar patterns (e.g., \citealt{kalnajs91,dehnen00,fux01,quillen05}), 
then we would expect those clumps and gaps
to only be seen in specific neighbourhoods where the resonant perturbations
are strongest, rather than throughout the disc.
In contrast, if clumps in the velocity distribution are primarily due to dissolution
of star clusters then they would
predominantly be comprised of homogeneous (in abundances and age) 
groups of stars \citep{eggen58a,freeman02,desilva07,bland10}.
The extent of structure in local velocity distributions outside 
the Solar neighbourhood is not yet known, though
dissection of test particle and N-body simulations suggests that
local velocity distributions 
could exhibit substructures such as gaps and clumps all over the 
Galaxy \citep{desimone04,quillen11,grand15}.

As the number of stars with accurate proper motions, radial velocities
and parallaxes increases with ongoing surveys, we can attempt
to break the degeneracies between models for the different dynamical mechanisms by studying variations
in the local velocity distributions as a function of position in the Galaxy or distance from the Sun.
The Radial Velocity Experiment  (RAVE; \citealt{rave}),  the
Large Sky Area Multi-Object Fibre Spectroscopic Telescope (LAMOST; \citealt{lamost}) survey,
The Geneva Copenhagen survey \citep{holmberg09}, 
the GAIA-ESO survey \citep{gilmore12,randich13},
the US Naval Observatory CCD Astrograph Catalog 5 (UCAC5; \citealt{zacharias17}), the 
Sloan Extension for Galactic Understanding and Exploration  survey (SEGUE; \citealt{yanny09}), 
and the Apache Point Observatory Galactic Evolution Experiment   (APOGEE; \citealt{majewski15})
are surveys that
have increased the number and distance of stars from the Sun 
with kinematic measurements compared
to the HIgh Precision Parallax COllecting Satellite   ({\it Hipparcos}) 
catalog \citep{perryman97} which primarily contains nearby
stars less than a hundred pc away \citep{dehnen98,famaey05}.  
The increased distances have made it possible to search for radial and vertical gradients
in mean velocities measured 
from local velocity distributions \citep{siebert11,williams13,carlin13,xu15,sun15,pearl17,carrillo18}.
One feature that is seen to vary as a function of position in the Galaxy
is that associated with the feature known as the 
{\it U-anomaly} or Hercules stream \citep{antoja14,monari17,perez17}.
Hyades and Sirius streams were identified in K giants up to 1.5 kpc from the Sun
\citep{wilson90}.

Fibre-fed spectroscopic 
surveys continue to expand on the number and accuracy of measurements of stars in the Galaxy.
Here we focus on kinematic and abundance measurements of half a million
stars from the GALAH survey.  Our focus is searching for variations
in substructure in local velocity distributions near the Sun.
From GALAH survey data we construct velocity distributions
from stars in angular regions defined by their viewed galactic longitude.
This gives us velocity distributions in local neighbourhoods that are displaced by a few 
hundred pc
from the the location of the Sun in the Galaxy.
In Section \ref{sec:vel}
we examine how velocity distributions vary as a function of viewed galactic longitude
and we use GALAH's spectroscopic measurements to see how
these velocity distributions depend on metallicity.
In Section \ref{sec:her} we examine the Hercules stream. 
In Section \ref{sec:near} we examine
clumps in the velocity distribution in nearby samples of GALAH stars to see
how low velocity streams vary with position near the Sun.

\section{GALAH spectroscopic survey of the Galaxy}
\label{sec:galah}

The GALactic Archaeology with 
HERMES (GALAH) survey\footnote{\url{http://galah-survey.org}} 
\citep{desilva15,martell17,kos17,kos18,duong18} 
is a Large Observing Program 
using the High Efficiency and Resolution Multi-Element Spectrograph 
(HERMES; \citealt{sheinis15}) 
with the 3.9m Anglo-Australian Telescope % (AAT) 
of the Australian Astronomical Observatory. % (AAO). 
Light is directed into HERMES from the 2dF multiple-fibre positioner \citep{lewis02}
so spectra of up to  392 stars can be obtained simultaneously.
GALAH is a high-resolution ($R = 28,000$) stellar spectroscopic survey of approximately a million stars,  
exploring the chemical and dynamical history of the Milky Way  \citep{desilva15}, that
measures radial velocities and stellar abundances for as many as 30 elements \citep{kos17}.

Spectroscopic measurements for the GALAH survey spectra are done with a data-driven approach,
using  the {\it Cannon},  
giving effective temperature, effective surface gravity  and 
abundance measurements from the stellar spectra 
(\citealt{ness15}; and see Section 2 by \citealt{sharma18}).
For more details on the GALAH data reduction pipeline and derived measurements
see \citet{kos17,martell17,sharma18}.
HERMES with 2dF and the same data reduction and measurement pipelines
are also used to characterize stellar targets for The Transiting Exoplanet Survey Satellite (TESS) 
(the TESS-HERMES program;  \citealt{sharma18}) 
and for spectroscopic follow-up of {\it Kepler} Satellite K2 fields 
(the K2-HERMES program; \citealt{witten18}). 

The GALAH input catalog stars are selected using near-infrared Two Micron All Sky Survey 
(2MASS; \citealt{skrutskie06})
JHK photometry of stars to estimate V band magnitudes using the conversion:
$V(J,K) = K + 2(J - K + 0.14) + 0.382\exp((J -K -0.2)/0.5)$ (following Section 5.1 by \citealt{desilva15}).
From these estimated V-band magnitudes, stars in the range 12--14 V mag are selected for observation. 
Stars are restricted to galactic latitudes  $|b|>10^\circ$
and declinations $-80^\circ < {\rm Dec} < +10^\circ$.  For more information about GALAH target 
selection, see Section 5.1 by \citet{desilva15,martell17}. 
The TESS-HERMES program target stars 
are also restricted in magnitude 
but with V band between 10 and 13.1 mag. As a result, that program
 contains brighter stars than the GALAH survey.
The K2-HERMES program extends to a fainter V mag of 15 and is restricted to redder stars with J-K$>0.5$.

The GALAH observed sample contains few previously known cluster stars (see Table 2 and
discussion by \citealt{kos18} and \citealt{martell17}), though open and globular clusters have been
observed with supplemental observational programs (K2-HERMES, the cluster and pilot programs
\citep{witten18,desilva18,duong18}).   
We attribute the lack of cluster stars in the GALAH survey
to logistics associated with fibre positioning (nearby targets on the sky cannot
be observed simultaneously) and scheduling of fields containing higher
densities of stars. 
%(fields in previously unobserved regions on the sky are often chosen before a field containing
%unobserved and previously observed target stars).

The spectroscopic parameters and abundances used here are those from GALAH 
internal data release 2 (iDR2) using {\it Cannon} version 2.7 \citep{buder18b},  
however Bayesian inferred estimates  
for distance and age are those from {\it Cannon} version 1.3 and as described by 
\citet{sharma18} in their Section 3.  
The differences in stellar parameters are minimal between the two data releases. 
The main difference is that more detailed abundances are available in iDR2 and for a larger sample
of stars. The number of stars in the data table that we are using is 539,594.

We use 
positions and proper motions from the  US Naval Observatory CCD Astrograph Catalog 5
(UCAC5) \citep{zacharias17}
where the precision level of the  Tycho-Gaia Astrometric Solution 
(TGAS) proper motions  \citep{michalik15} is extended to many millions more stars. 

The Gaia-ESO Survey  \citep{gilmore12,randich13} contains a sample of about 
100,000 stars over 14 -- 19 in V band, 
most of which are thick disc and halo stars, as well as specially targeted star clusters. 
In contrast GALAH has more stars, a narrower magnitude range, lacks star clusters
and contains more dwarf and thin disc stars.
The near-infrared APOGEE survey targeted the plane of the disc with a sample of 150,000 red giants. 
As the survey targets giants, APOGEE stars tend to be more distant than GALAH stars.
The APOGEE-S survey is carrying out a survey of similar design and scale from the southern hemisphere.
The magnitude ranges and different regions of the Galaxy covered make 
both Gaia-ESO and APOGEE are complementary to the GALAH sample, with minimal overlap. 
The RAVE survey \citep{rave}  contains a similar number of stars as the current GALAH sample and there
is some overlap with GALAH. 
The RAVE sample spanning magnitudes from 8 -- 12 in the I band were selected from multiple input sources over the time span of the survey and includes a colour selection of J-K mag $> 0.5$ 
so it preferentially selects giants 
(see \citealt{wojno17}). In contrast, the GALAH sample is sourced solely from 2MASS photometry, has a well defined selection function, and when compared to RAVE, is dominated by local disc dwarfs making it more sensitive to nearby substructure in the Galaxy. 

\subsection{Coordinates}
\label{sec:coord}

To discuss the velocity distribution of stars in the Galactic disc we  
first define a Galactic coordinate system.
We adopt a Cartesian galactocentric coordinate system with 
origin at the Galactic centre and the Sun at $(x_\odot, y_\odot, z_\odot) = (-R_\odot, 0,0)$,
where $R_\odot$ is the galactocentric radius of the Sun.
We  ignore the offset of the Sun from the Galactic plane 
($z_\odot = 25 \pm 5$ pc, \citealt{bland16}).
In a cylindrical coordinate system ($r,\theta,z$), the velocity of a star has radial, tangential and
vertical components ($v_r,v_\theta,v_z$).   The local standard of rest (LSR) corresponds to a
circular orbit at a radius $R_\odot$, in the Galactic plane,
with $v_r= v_z = 0$ and $v_\theta = -V_{circ,\odot}$.
The velocity
$V_{circ,\odot}$ is the speed of a star in a circular orbit  at the Sun's radius (ignoring
non-axisymmetric perturbations by spiral arms or the Galactic bar).  

By a common convention,
the velocity components $(u,v,w) = (-v_r, -v_\theta - V_{circ,\odot}, v_z)$ 
are velocities in cylindrical coordinates
subtracted from the local standard of rest, with $u$ equal to minus 1 times the radial velocity component,
and with large positive $v$ for stars at  high angular momentum coming from the outer Galaxy
into the solar neighbourhood.  The negative sign for $v_\theta$ gives a Cartesian coordinate system
for Galactic rotation with positive $z$ toward the north Galactic pole.
 
In the inertial Galactic coordinate system, we adopt space motions for the Sun 
$v_{\theta,\odot} =-242.0$ \kms, $v_{r,\odot} =11.1$ \kms, and $v_{z,\odot}=7.25$ \kms.
The galactocentric radius of the Sun $R_\odot =8.0$ kpc is that by \citet{reid93},
see  \citet{sharma14}, Sec 2.5, for discussion regarding this choice.
The radial and vertical peculiar space motions of the Sun (with respect to LSR) 
$v_{r,\odot}$  and $v_{z,\odot}$ are by \citet{schonrich10}.
The tangential space motion of the Sun
$-v_{\theta,\odot}=V_{circ,\odot} + v_{\odot,pec}$, is observationally well constrained by 
the angular rotation rate of Sgr A*  % $\Omega_{SagA*} = 
(30.24  \kmsikpcns; \citealt{reid04}; see \citealt{sharma14}, Sec 6.3).
For $R_\odot=8.0$ kpc, the angular rotation rate
of Sgr A*  gives a tangential velocity $-v_{\theta,\odot} = 30.24 \times 8.0 = 241.92$ \kmsns,
consistent with the adopted value given above and used to compute $v_\theta$ for each star.
%This is free of controversy about  $V_{circ}$ and $V_{\odot,pec}$  
% and hence my favourite value for the Sun's tangential velocity.
Here $v_{\odot,pec}$ is the Sun's peculiar velocity with respect to the LSR. 
For $v_{\odot,pec} \approx 10$ \kms (as found by \citealt{schonrich10}) the circular velocity 
at the galactocentric radius of the Sun
$V_{circ,\odot} \approx 230 $ \kms (see Section 5.3.3 by \citealt{bland16}).

From positions, proper motions and distances, GALAH spectroscopically measured radial velocities 
and using the Sun's space motion (as described above)
the galactocentric cylindrical coordinates are computed for each star in the survey
$[\theta,r,z, v_\theta,v_r, v_z]$.  
From these and using $V_{circ,\odot} = 230 $ \kms we compute velocity components ($u,v,w$).
Our choice for $V_{circ,\odot}$ is 30 \kms higher than used by \citet{dehnen98} 
(who adopted $V_{circ,\odot} = 200 $ \kms, 
a somewhat lower $v_{\odot,pec} =  5.2 $ \kms for the Sun and the same $R_\odot = 8 $ kpc).
As a consequence our $v$ values for clumps identified in the velocity distributions are consistent with but
20--30 \kms lower than those reported by \citet{dehnen98}. 
%
%I think \citet{liang17} is using same convention as Dehnen.
%There is discussion by \citet{famaey05} but pretty confusing on what has been subtracted from what.

%\citet{antoja14} likes this
%$ V_{circ,\odot} = 238$ \kms following recent results by Honma et al. (2012) 
% based on VLBI astrometry of Galactic maser sources. 

\section{Velocity distributions constructed from stars at different galactic longitudes}
\label{sec:vel}

To study local velocity distributions we  construct
histograms in radial and tangential velocity bins that are 4 km/s square. 
The numbers of stars in each bin gives the distribution of stars
as a function of the two in-plane velocity components.
We construct velocity distributions from stars in different angular regions in galactic longitude.
Central galactic longitudes for each region range 
between $\ell=180^\circ$  (the Galactic
anticentre direction) and $\ell=40^\circ$, in steps of $20^\circ$,
 and covering the Galactic centre direction at $\ell = 0^\circ$.
We neglect the region between $\ell=40^\circ$ and $\ell=180^\circ$.  
It is not well sampled because the GALAH survey has few stars in this direction on the sky.  
Each region is 
$\Delta \ell = 40^\circ$ wide in galactic longitude, so the neighbourhoods we use
to construct histograms overlap each other.   For example, the histogram centered
at $\ell = 260^\circ$ is constructed from stars with $240^\circ < \ell < 280^\circ$.
We eliminate stars with high or low galactic latitudes, $|b|>45^\circ$, and
stars with distance errors $\sigma_d/d > 0.5$, where $d$ is the distance and $\sigma_d$
its computed error (standard deviation; see Section 3 by \citealt{sharma18}). 
For stars with $d<1$ kpc, 500 pc, or 250 pc, the median values of $\sigma_d/d$ are 0.23, 0.18 and 0.14.
Table \ref{tab:cuts} lists the cuts made to the survey stars relevant for each Figure and Table
in this manuscript.

We also eliminate stars observed as part of 
dedicated observing programs of open clusters \citep{desilva18}, and GALAH pilot surveys \citep{duong18}, 
however we did not remove stars observed as part of the K2-HERMES \citep{witten18}
or TESS-HERMES \citep{sharma18} observing
programs because they increase the number of stars observed near the Galactic anticentre.
We have checked that removal of the K2-HERMES and TESS-HERMES program
stars does not impact the appearance of the histograms.
After excluding those with large distance errors, that are
part of open cluster observing programs or have $|b|>45^\circ$, 
the number of GALAH survey stars remaining in iDR2 is 376,510.
Within this group, the number of stars observed as part of 
K2-HERMES and TESS-HERMES programs is 69,162.

Each histogram is  constructed from stars within a specific metallicity range and a specific
range in estimated distance.
Because each histogram is constructed from a distance limited sample and only along a particular
viewing direction, the median position in galactic ($x,y$) coordinates of the stars
is not at the position of the Sun.  Thus each histogram gives a velocity distribution
constructed from stars in different local neighbourhoods.

\begin{figure*}
\includegraphics[width=18.5cm,trim={5mm 0 0mm 0},clip]{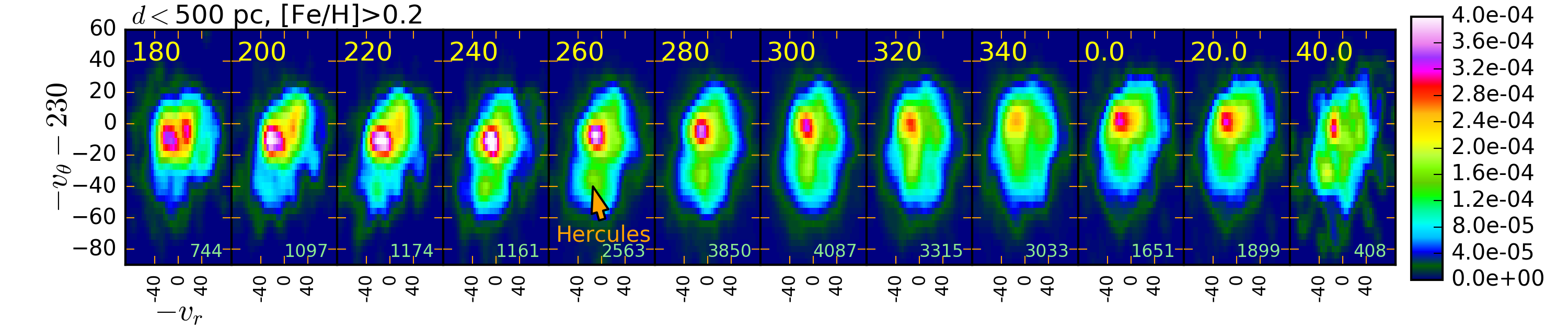}
\includegraphics[width=18.5cm,trim={5mm 0 0mm 0},clip]{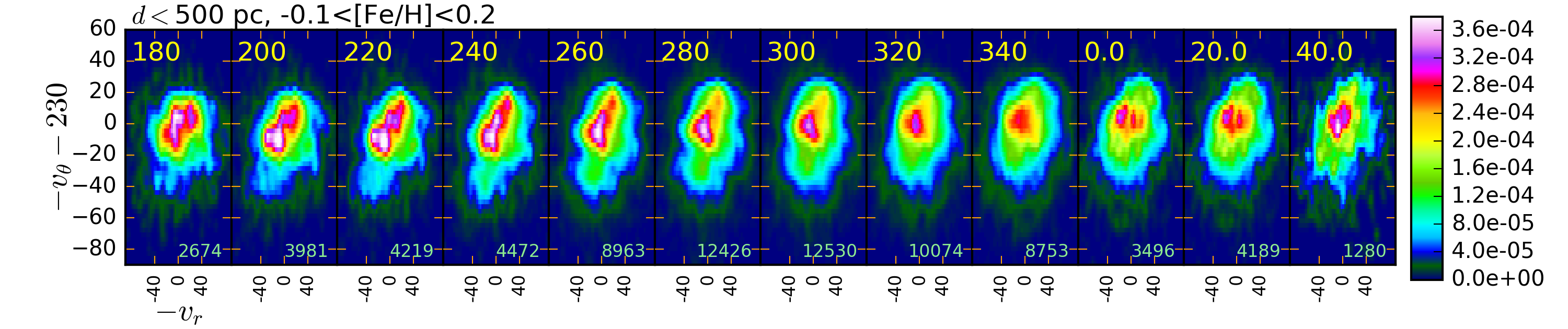}
\includegraphics[width=18.5cm,trim={5mm 0 0mm 0},clip]{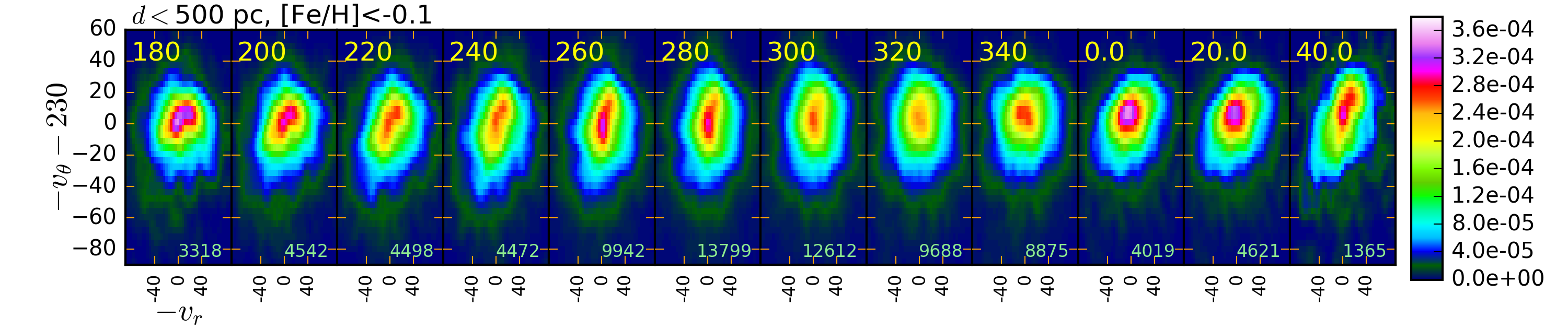}
\includegraphics[width=18.5cm,trim={5mm 0 0mm 0},clip]{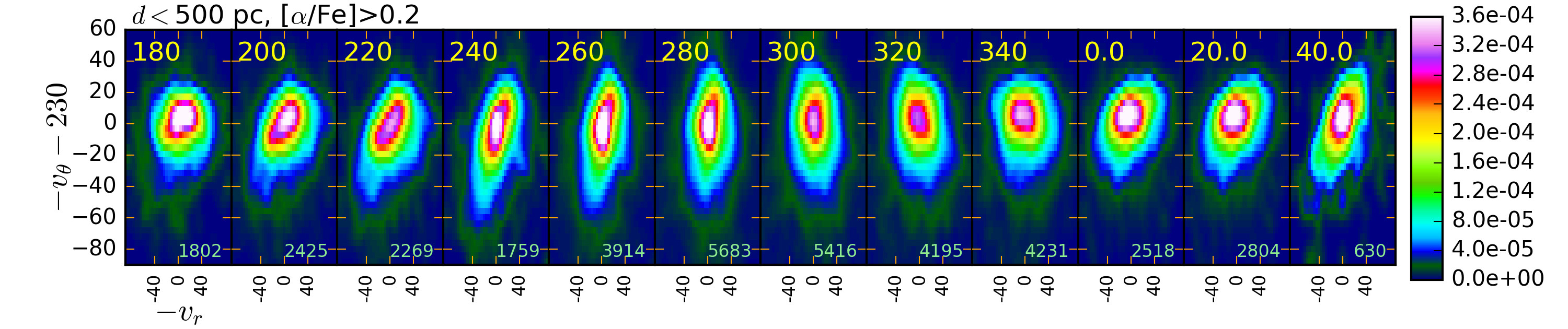}
\caption{Histograms showing $uv$ velocity distributions constructed of stars viewed at 
different galactic longitudes.  
The x-axes show the radial velocities $u = -v_r$  of each bin in \kmsns.
The y axes show the tangential velocities $v = -v_\theta - 230 $ \kms of each bin in \kmsns.
The histograms only contain nearby stars with distances  $d<500$ pc. 
The central galactic longitude for each histogram is the yellow number on the top of each histogram
and the number of stars used to construct the histogram shown on the bottom.
Each histogram is created using stars within $20^\circ$ of the central galactic longitude.
Each row of histograms is for stars within a different metallicity range.
From top panel to bottom panel 
a)  [Fe/H] $ > 0.2$.
b) $-0.1 < $ [Fe/H] $ < 0.2$.
c) [Fe/H] $ < -0.1$.
d)  [$\alpha$/Fe] $ >0.2$. 
\label{fig:line}}
\end{figure*}

Figure \ref{fig:line} shows 4 sets of histograms, each constructed only of stars within $d<500$ pc.
Figure \ref{fig:line}a (top panel) shows metal rich stars with [Fe/H] $ > 0.2$, 
Figure \ref{fig:line}b (second from top panel) shows near solar metallicity stars with $-0.1<$ [Fe/H] $ < 0.2$,
Figure \ref{fig:line}c (third from top panel) show lower metallicity stars with [Fe/H] $ < -0.1$
and  Figure \ref{fig:line}d (bottom panel) show high  
$\alpha$-element (thick disc) stars with [$\alpha$/Fe] $ > 0.2$.
The distance limit eliminates giants due to the magnitude limit of the GALAH survey
and we have checked that removal of stars with stellar
surface gravity $\log g < 3.5$ does not affect the appearance of the distributions.
Each histogram has 
x-axis showing the radial velocities $u = -v_r$ of the velocity bins and 
y-axis the tangential velocities $v = -v_\theta - 230 $ \kms of the velocity bins.
Each panel in a row is for stars restricted to within 20$^\circ$ of a particular galactic longitude
with central galactic longitude beginning at $\ell=180^\circ$, on the left, and increasing in $20^\circ$
steps to the right.
The central galactic longitudes are labelled on the top of each histogram and the number
of stars used to make each histogram is labelled on the bottom.

The histograms are normalized so that they integrate to 1 and are displayed with the same colourbar.
The numbers on the colourbar show normalized counts in each bin.
Specifically the numbers shown on the colourbar are the number of stars in the bin divided by
the bin area (16 (km~s$^{-1}$)$^2$) divided by the number of stars in the histogram.
The number of stars in each histogram impacts the precision of the velocity distribution.
Histograms with few stars exhibit clumps that are due solely to the small numbers of stars
in each bin.
For reference, the panel of Figure \ref{fig:line}a at $\ell=200^\circ$ has about 1000 stars. Its
peak 4 \kms square  bin 
contains 12 stars and has a value of 0.0007 in the normalized histogram (and corresponding
to the numbers shown on the colourbar).    In the same histogram, the dark green contours
have only a couple of stars per bin.

%0.0003 has 6 stars of 1000
%Take the number and multiply by 1000 you get 0.7
%Approximately take number on colourbar multiply by the number of stars to get numbers in each bin.

We consider how measurement errors in radial velocity and proper motions
affect the ($u,v$) errors at different galactic longitudes.
Because we have discarded high and low galactic latitude stars we can 
neglect galactic latitude for this
discussion.
At a viewing angle with galactic longitude $\ell = 270^\circ$,
 the $v$ velocity component
is determined by the radial or line of sight velocity
and the $u$  velocity component
is determined by the proper motions.  At a galactic longitude of $\ell = 0$ or $180^\circ$
the opposite is true; the $v$ velocity component
is determined by the proper motions and the $u$ velocity component
is set by the radial velocity.   Thus the errors in ($u,v$) form an ellipse that is dependent
on viewing direction.

Line of sight (radial) velocity component errors (standard deviations)
from the GALAH data reduction pipeline 
are currently about  $\sigma_{vlos} \approx 1$ \kms   (see \citealt{kos17}
and Figure 6 by \citealt{sharma18}).\footnote{Ongoing improvements in the data reduction
pipeline are expected to decrease the error to $\sim 0.1$\kms, Janez Kos, private communication.}
The tangential velocity component of a star $v_{tan} =  \mu\, d$ where $d$ is
the distance from the Sun and $\mu$ is the proper motion.
For each star with error in distance ratio $\sigma_d/d < 0.5$ (as those are used to make
our histograms) 
we estimate the error in the tangential velocity component $\sigma_{vtan}$ 
from the distance $d$,  distance error $\sigma_d$,  
the UCAC5 proper motion $\mu$ and its error $\sigma_\mu$,
\begin{equation}
\left( \frac{\sigma_{vtan}}{v_{tan}}\right)^2 = 
\left( \frac{\sigma_d}{d} \right)^2 + 
\left( \frac{\sigma_\mu}{\mu} \right)^2 .
\end{equation}
We find a median error in the tangential velocity component of 
$ \sigma_{vtan} \approx  $ 1.2  and 1.8  \kms  for stars 
with distances within 0.5 and 1 kpc,
and 2.7 \kms for stars at all distances.  As the error for the nearer stars
is similar to that of the line 
of sight velocity component, $ \sigma_{vtan} \sim \sigma_{vlos}$
there is no strong directional dependence 
(dependence on galactic longitude) to
the $u$ and $v$ velocity component errors.
The size of the estimated tangential velocity errors are small enough that
they would not have obscured structure smaller than a few \kms
in our histograms.
Thus variations in the numbers of clumps in the velocity distributions with galactic longitude
is not caused by a difference in the sizes of errors in $u$ compared to $v$
that is sensitive to direction on the sky.
%Even though the tangential velocity errors are smaller
%than the radial velocity errors, the fewer
%clumps in the velocity  distribution seen near in the direction of the Galactic centre
%($\ell = 0^\circ$)
%compared to near the anticentre or to $\ell = 270^\circ$

Some stars in iDR2 are binaries and their radial velocities may include
binary reflex motions (e.g., \citealt{elbadry18,badenes18}).
Using multiple epoch APOGEE observations, \citet{badenes18} 
found that 5.5\% of main sequence or subgiant stars (see their Table 1) 
exhibited a change in measured
radial velocity (between epochs)
greater than 10 \kms.  Binaries would blur or smooth the velocity distributions,
adding faint wings onto high density peaks.   Our histograms are essentially
an underlying velocity distribution convolved or smoothed with an error function 
that depends on 
 the distribution of binary star motions.
However, this blurring should not obscure strong peaks in the velocity distribution 
as only 5--10\% of the stars are affected (with  $\sim$ 10\% of stars 
with variation in radial velocity greater than 1 \kms, estimated assuming
a similar fraction of binary stars per decade in log semi-major axis).
Though they may be included in GALAH DR2, 
repeat observations (spectra taken at different epochs)
of GALAH survey stars are not present in our data table.

A clear chemical separation in the [$\alpha$/Fe] vs [Fe/H] abundance plane can be made between 
thin and thick disc populations 
(see for example \citealt{bensby07b,adibekyan12,bensby14,hayden15}).
The {\it Cannon}'s [$\alpha$/Fe]  %is trained on the one from SME, which
is an error-weighted mean of O, Mg, Si, Ca, and Ti abundance measurements \citep{buder18a,buder18b}.  
Because the Ti measurements
are the most precise, the [$\alpha$/Fe] measurement is most strongly influenced by the Ti abundance.
We chose [$\alpha$/Fe] $>0.2$  for 
Figure \ref{fig:line}d so as  to exclude the 
$\alpha$-element poor or thin disc stellar population, leaving
 primarily stars from the older $\alpha$-element rich and thicker disc. 
See \citet{fuhrmann98,bensby03,bensby14,duong18} 
for discussion on the divisions between the two populations.
The current GALAH pipeline mis-classifies some cool giants as dwarfs, giving a spurious 
excess of nearby high $\alpha$ stars in the distance distribution.   To minimize this contribution 
we also removed stars with effective temperature below $4500^\circ$ to make 
the bottom panel of Figure \ref{fig:line}.
The distance distribution used to make the velocity distribution of Figure \ref{fig:line} bottom panel
is shown in section \ref{sec:dist}.
The median error in [$\alpha$/Fe]  
reported by the {\it Cannon} software
for the $d<500$ pc stars is 0.06.  
% actually was lower, but I did not believe it   
%(where for each star we have added the errors for [Mg/H]  and [Ti/H]
 %in quadrature to compute its error  in [$\alpha$/Fe] 
%then computed the median of all stars we used to make our histograms).
The median error is similar for the [Fe/H] abundance;  the reported error is 0.054 according
to Figure 6 by \citet{sharma18} comparing the {\it Cannon} measurements to those in the training set.

We estimate the location of the neighbourhoods of our histograms based on the mean distances
of the stars in each neighbourhood from the Sun. 
The median distances for the $d<500$ pc sets are 
$d_n = $ 340, 330, 330 and 320 pc (from the Sun) 
for the metal rich, solar, metal poor and high-$\alpha$ groups of stars, respectively, 
shown in the histograms 
of Figure \ref{fig:line}. See appendix \ref{sec:dist} for complete distributions of distances 
for stars in each metallicity range. 
We have checked histograms in bins of $v$ and $d$ made for stars seen in different galactic longitudes
and sets of histograms  in bins of $u$ and $d$.  Neither set showed 
sensitivity of the distance distributions on direction, 
though we re-examine this issue in Section
\ref{subsec:hem} when we compare velocity distributions from stars in the northern Galactic
hemisphere to that in the south.

% fig1

\section{The Hercules stream}
\label{sec:her}

The ($u,v$) velocity distributions
for metal rich stars shown in Figure \ref{fig:line}a  exhibits an extension to lower
tangential velocities $v$ at about
$v \sim -45 $ \kms known as the Hercules stream.    The stream is annotated in
the galactic longitude $l=260^\circ$ histogram with an orange arrow in Figure \ref{fig:line}a (top panel).
The Hercules stream (also called the U-anomaly) is one of the most prominent streams 
or clumps seen in the velocity distribution of stars in
the solar neighbourhood  \citep{eggen58b,eggen65c,dehnen98,famaey05,ramya16}.   
The Hercules stream is seen as a separate feature
because it is separated by a dip or gap in the numbers
of stars  at $v \sim -30$ \kms 
from stars in more nearly circular orbits at $v \sim 0$.
The stream is centered at a negative radial velocity component $u\sim -20$ \kms corresponding to stars
moving away from the Galactic centre.

Stars in the Hercules stream
span a range of ages and metallicities (e.g. \citealt{dehnen98,famaey05,bensby07,bovy10,ramya16}) 
implying that the division in phase space 
has a dynamical origin.  The Hercules stream is thought to be caused by perturbations
from the Galactic bar \citep{dehnen99b,dehnen00,fux01,minchev10,antoja14,portail17a,perez17,monari17}.
For a short and quickly rotating or {\it fast} bar, the Hercules stream is related to 
the bar's outer Lindblad resonance  \citep{kalnajs91,dehnen00,minchev07,antoja14,monari17,hunt18}. 
Stars in nearly circular orbits are outside this resonance and have orbits preferentially
elongated parallel to the bar.
Stars in the Hercules stream are inside resonance and in some models they
are elongated perpendicular to the bar \citep{kalnajs91,dehnen00,fux01,minchev10}, 
with two radial epicyclic oscillations per orbit in the bar's rotating frame.
For a longer and slower bar, the stream is caused by
orbits associated with a stable Lagrange point in the bar's corotation region 
 \citep{portail17a,perez17}.
The galactocentric radius of corotation is that where the bar angular rotation rate or pattern speed is 
equal to the orbital angular rotation rate of a star in a circular orbit in a galaxy with the same azimuthally 
averaged mass distribution and lacking the bar.
A fast rotating bar has  
pattern speed $\Omega_b= (1.89 \pm 0.08) \Omega_\odot$
or $\Omega_b=56 \pm 2$ \kmsikpc  for a solar angular rotation rate
$\Omega_\odot = 29.5 $ \kmsikpc  \citep{antoja14}. 
For the {\it slow} bar $\Omega_b=39 \pm 2 $  \kmsikpc   \citep{perez17,portail17a}.
N-body simulations show that spiral structure may influence the appearance of the Hercules stream
in velocity distributions  \citep{quillen11,grand15}.

Recent studies have shown that
the visibility or contrast of the Hercules stream in the ($u,v$) or ($v_r,v_\theta$)  velocity distribution 
of stars in the solar neighbourhood is dependent on 
metallicity \citep{liu15,liu16,perez17}, with higher metallicity stars more clearly showing the 
gap separating the stream
from the more nearly circular orbits at low $|u|,|v|$ (see Figure 5 by \citealt{perez17}).
The stream is predicted to be most easily detected in the velocity distribution 
for giant stars that are within 3 kpc  from the Sun and observed
near a galactic longitude of approximately $270^\circ$ \citep{hunt18}.
We confirm both of these trends. Figure \ref{fig:line}a (top panel) 
shows that the stream is most prominent
at galactic longitudes $220^\circ < \ell < 340^\circ$ but is also weakly seen at $\ell \sim 40^\circ$.
A comparison of Figure \ref{fig:line}a (top panel), for higher metallicity stars with [Fe/H]$>0.2 $, 
with Figure \ref{fig:line}c (third from top), for lower metallicity stars with [Fe/H]$<-0.1 $,
shows that the Hercules stream is predominantly seen in higher metallicity stars.
The Hercules stream is also visible in the solar metallicity stars  in 
Figure \ref{fig:line}b (second from top)  corresponding to stars with  $-0.1 <$ [Fe/H] $<0.2 $
confirming   \citet{raboud98,liu16,ramya16,perez17} 
who find the Hercules streams consists of intermediate to high metallicity stars. 
The stream is hardest to detect in the low metallicity stars with  [Fe/H] $<-0.1$ shown 
in Figure \ref{fig:line}c (third from top).

\begin{figure}
\includegraphics[width=8.0cm]{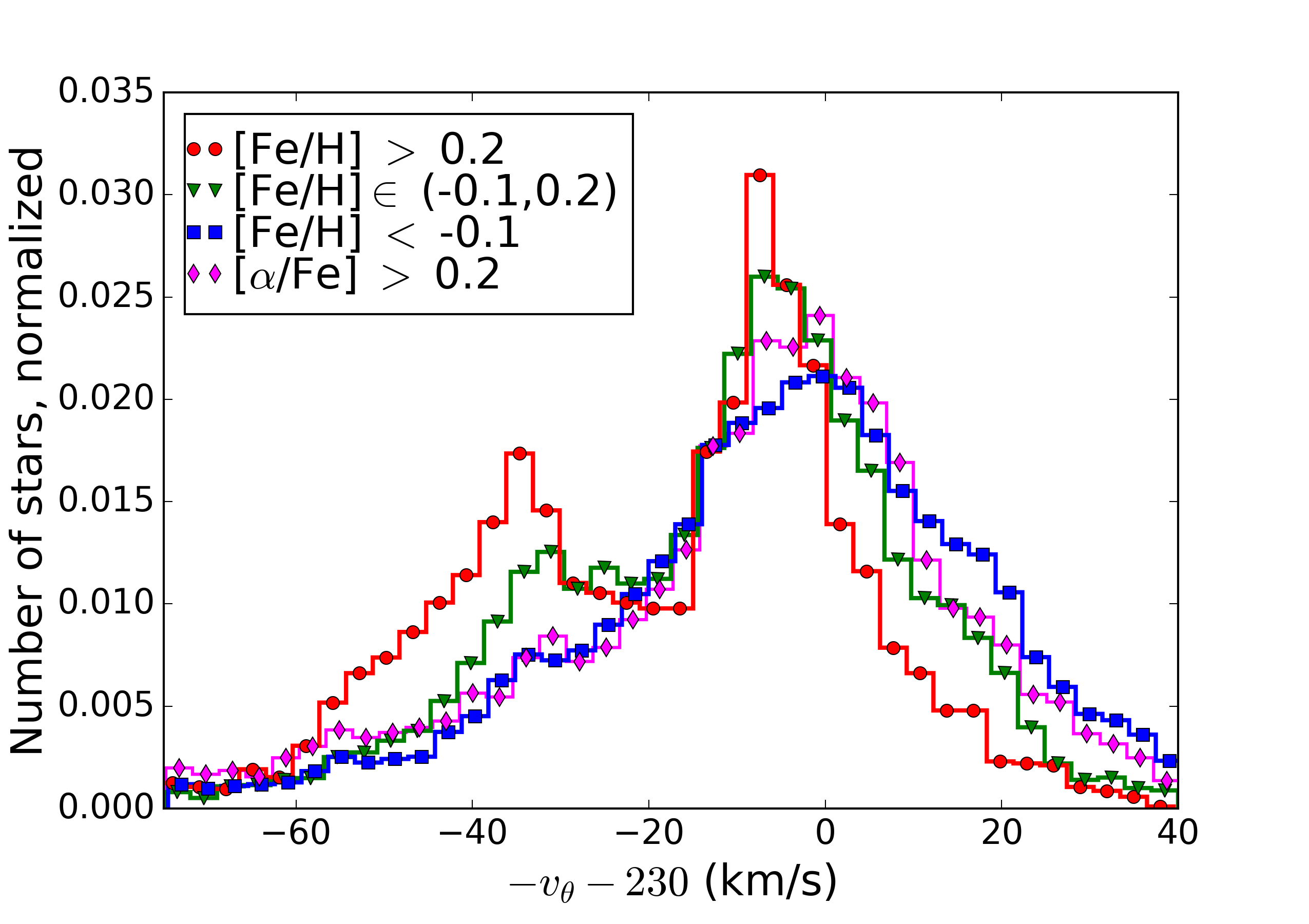}
\caption{The distribution of tangential velocities for stars 
restricted in galactic longitude to
 $250^\circ < \ell < 290^\circ$, in radial velocity to 
 $-40 <  -v_r < 10$ \kms and in distance to less than 1 kpc.
The Hercules stream is the peak at 
a tangential velocity $v = -v_\theta - 230 \approx 35 $\kms
that is prominent in the supersolar metallicity stars shown in red and with round points.
There are more subsolar metallicity stars (shown in blue with square points) 
at positive tangential velocities, corresponding to stars
coming into the solar neighbourhood from the outer Galaxy.
Near solar metallicity stars are shown in green with triangles
and $\alpha$-element enhanced  stars are shown in magenta with diamonds.
\label{fig:vhist}}
\end{figure}

To illustrate how the visibility of the Hercules stream depends on metallicity, we show 
vertical slices near longitude $\ell = 270^\circ$ through histograms like those shown in Figure \ref{fig:line}.
In Figure \ref{fig:vhist} we plot one dimensional histograms showing the tangential velocity distributions
 for stars in the four different metallicity ranges; [Fe/H]$>0.2$, $-0.1<$[Fe/H] $<0.2$, [Fe/H]$<-0.1$
 and $[\alpha$/Fe] $>0.2$.
 Stars are restricted in galactic longitude
to $250^\circ < \ell < 290^\circ$, in radial velocity to $-40 <  -v_r < 10$ \kms, 
in distance to less than 1 kpc and Galactic latitude $|b| < 45^\circ$.
The histograms have been normalized so that they integrate to 1 and the bins are 3 \kms wide.
In each histogram the number of stars is 2700,   8982, 11339, and 
4114 for the metal-rich, near-solar, metal-poor and high-$\alpha$ metallicity stars.

The Hercules stream is the red peak in Figure \ref{fig:vhist}  at 
a tangential velocity $v = -v_\theta - 230 \approx -35 $\kms and it is clear because
of the dip at $v \approx -25$ \kms separating
it from stars in more nearly circular orbits at $v \sim 0$ \kmsns.
A dip and peak are visible in metal-rich and near-solar metallicity stars.
As there are more  metal poor stars in these histograms than metal-rich stars
the absence of a dip  in the histogram at $v \sim -25 $\kms in the metal
poor sample is significant.  
We disagree with the previous studies by \citet{bensby07,bovy10} 
who find no significant metallicity preference 
in the Hercules stream for stars in the solar neighbourhood.

There are similar numbers of $\alpha$-element enhanced  stars
as supersolar metallicity stars in the histograms of Figure \ref{fig:vhist}, 
but many of the $\alpha$-element enhanced  stars appear to have nearly circular orbits,
suggesting that there might be  contamination from low temperature
dwarfs that have less well measured abundance ratios. 
The smaller number of stars in the histograms  of
the $\alpha$-element enhanced stars with negative $v$ velocities,
makes each velocity bin in that region more poorly sampled. 
Figure \ref{fig:line}d (bottom panel), illustrating velocity distributions for 
the high $\alpha$-element stars at different longitudes, 
shows that these also contain an extension to low $v$ that
is most strongly seen at galactic longitudes $240 < \ell < 280^\circ$. This extension
reaches somewhat lower tangential velocities than the Hercules stream seen in 
high metallicity stars (those with [Fe/H]$>0.2$ and shown in Figure \ref{fig:line}a; top panel).
We do not see a gap in  Figure \ref{fig:line}d or a dip in Figure \ref{fig:vhist} in the high-$\alpha$ stars.
A dip must be present in the velocity distribution for us to identify a peak with
the Hercules stream.
We would require better measurements and sampling of the $\alpha$-rich population to 
determine whether the Hercules stream really is absent from that
stellar population.

The lack of low metallicity stars in the Hercules stream  imply that the 
the Hercules stream is predominantly
a metal-rich and thin disc dynamical structure.  We support the similar conclusion
by \citet{ramya16}; 
for a plot of [$\alpha$/Fe] vs [Fe/H] for giant stars chosen specifically in
the Hercules stream see their Figure 7. 
Because of the Galaxy's metallicity gradient, higher metallicity thin disc stars are
preferentially found closer to the Galactic centre, at smaller galactic radius.
Stars in the Hercules stream with negative $v$ have lower angular momentum
than a circular orbit at the solar radius.   Using a circular velocity of 230 \kms
a star with $v = -40 $ \kms and currently at radius $r=R_\odot$ 
would have a mean orbital or guiding radius of $r_g = (230-40)/230 \times 8 =  6.6$ kpc
or 1.4 kpc closer to the Galactic centre than the Sun.
Using a metallicity gradient of -0.07 dex per kpc in [Fe/H] \citep{anders17},  stars 
at this guiding radius would on average have a metallicity of 
[Fe/H] $\sim 0.1$, where we are taking a solar value of [Fe/H] $=0 $  at $R_\odot$
(e.g., \citealt{casagrande11}).
This is only 0.1 dex away from our cut-off of 0.2 used to make the histograms shown
in Figure \ref{fig:line}a.   At a guiding radius of 6.6 kpc, there would be fewer lower
metallicity stars than at the Sun's guiding radius.  The lack of metal poor disc
stars at this guiding radius likely accounts for the dominance of the Hercules
stream in the higher metallicity stars compared to the lower metallicity stars.

In this discussion we have so far neglected the role of the Galactic bar, but it is 
necessary to account for the gap separating the Hercules stream from the lower
eccentricity stars and account for the non zero $u$ velocity of the stream (e.g., \citealt{dehnen99b}).
The gap is attributed to a division in phase space caused by one or more resonances with
the bar \citep{kalnajs91,dehnen00,fux01,perez17}.   Orbits can be classified based on
their proximity to a periodic orbit or equivalently an orbit that is closed in a frame rotating with a bar.
For the fast bar model, the simplest orbit families
are those just outside the outer Lindblad resonance and elongated parallel to the bar
and those just inside the inner Lindblad resonance that are elongated perpendicular to the
bar  \citep{kalnajs91,dehnen00,fux01}.   
The gap seen in the velocity distribution could be associated with an unstable or chaotic
region separating these two orbit families.  More complex models (e.g. \citealt{fux01,minchev10})
find that additional orbit families are present so it is not straightforward
to associate each velocity vector (or ($u,v$) components) with a particular orbit shape.   
For the slow bar model,  the gap in velocity space is associated with corotation 
and separates orbits
circulating around the Galactic centre from those librating about an L4 or L5 Lagrange point
(see Figure 3 by \citealt{perez17}).
  
The high $\alpha$-element or thick disc population has a shorter radial
exponential scale length and a higher
velocity dispersion than the thin or low $\alpha$-element disc \citep{hayden15,hayden17}.
A higher velocity dispersion implies there are more stars at higher eccentricities that may
be affected by resonances with the bar that can push them into the solar neighbourhood.
The extension to low $v$ and negative $u$ in the velocity distributions of the 
high $\alpha$-element stars
would influence measurements of velocity component averages, 
and is relevant for interpretation of 
non-circular streaming motions seen in the thick
disk RAVE survey stars  \citep{antoja15}. 

\subsection{The gap between the Hercules stream and other stars}
\label{sec:gap}

%How nearby are these apogee giant stars?  answer: 1-4 kpc
%Using Radial Velocity Experiment  (RAVE; \citealt{rave}) and 
%Large Sky Area Multi-Object Fibre Spectroscopic Telescope (LAMOST; \citealt{lamost}) survey data,

Using RAVE and  LAMOST survey data,
\citet{perez17} 
found that the $v$ value of the gap  separating the Hercules stream
from stars on circular orbits in the velocity distribution depends on distance from the Galactic centre, with
variations seen over a distance of a few hundred pc.  
This sensitivity of the local velocity distribution to galactic position
was also seen in their numerical model.
A similar shift in $v$ as a function of galactocentric radius was seen by \citet{antoja14} in RAVE
survey data and predicted with a fast bar model.
For  stars  at a galactic radius of 8.6 kpc \citet{perez17} saw the gap at $v \approx -50$ \kmsns,
whereas at a radius of 7.8 kpc (with their adopted  solar radius of 8.2 kpc),
the gap was at $v \approx -30$ \kmsns.  Thus they found that the gap $v$ velocity 
increased (was less negative) with decreasing galactic radius.
%They adopted a solar radius of 8.2 kpc.

\begin{figure*}
\includegraphics[width=10.0cm]{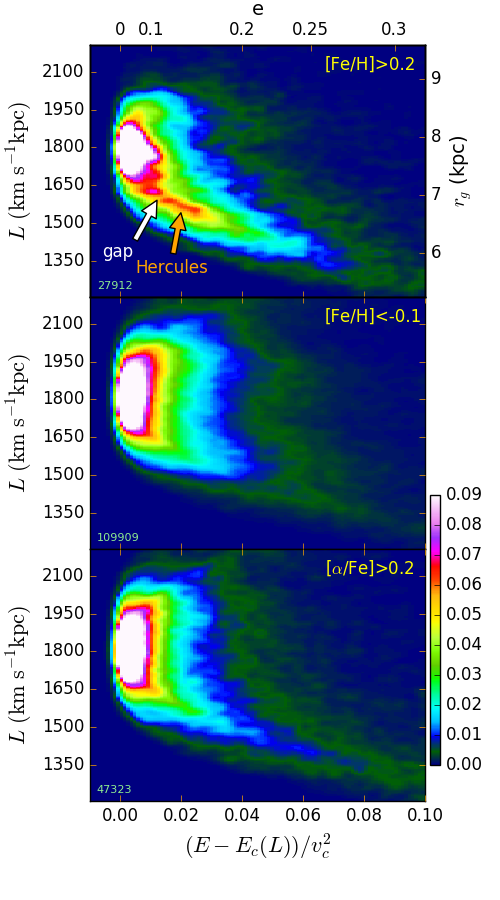}
\caption{Histogram of stars in angular momentum and energy  bins.
The x-axes are the difference between orbital energy in the plane
and that of a circular orbit normalized by $v_c^2$.  
The y-axes are angular momentum (the z-component) in units of \kmskpcns.
%All three histograms are for stars with distances $d<1$ kpc.
The top panel shows higher metallicity stars with 
Fe/H] $> 0.2$, the middle one is for lower metallicity stars with  [Fe/H] $<-0.1$
and the bottom one is for high $\alpha$-element stars with [$\alpha$/Fe] $> 0.2$. 
Green numbers in each panel on the lower left are the number of stars used to make the histogram.
The high metallicity and high $\alpha$-element stars show extensions to low
angular momentum and higher energy, corresponding to higher eccentricity
stars from the inner Galaxy reaching into the solar neighbourhood.
There are more  high angular momentum low metallicity stars corresponding to stars from the outer
Galaxy coming into the solar neighbourhood.
There is a gap at about 1600 \kmskpc separating the Hercules stream
from the more circular orbits.  In the topmost histogram the right axis
shows guiding radius computed from the angular momentum and assuming
a flat rotation curve.  The top axis shows orbital eccentricity, computed with 
 energy difference equal to the square of the eccentricity and for a flat rotation curve.
%  (see appendix\ref{sec:dyn}).  
\label{fig:LE}}
\end{figure*}
%fig2

An orbital resonance is defined by a commensurability between orbital frequencies.
Equivalently there is an integer relation between 
orbital and epicyclic periods and the period of a perturbation, here the Galactic bar.
In the Galactic disc, the orbital period is approximately set by the angular momentum.
So a division in phase space at a particular angular momentum could be associated
with a resonance with the bar.
The dependence of the velocity distribution on the radius of the neighbourhood 
suggests that the Hercules stream or the gap separating it from circular orbits is set by a particular
angular momentum value.   
To test this hypothesis we use the GALAH stars to make histograms of angular momentum versus a
normalized
energy difference $(E - E_c(L))/v_c^2$ that is approximately equal to the square of orbital eccentricity
(see equation \ref{eqn:energy_ee}). 
The energy $E$ is the energy per unit mass restricted to velocity components in the plane (defined
in equation \ref{eqn:EE}), and $E_c(L)$ (equation \ref{eqn:EL})  is the energy per unit mass
of a circular orbit with angular momentum $L$.
The $z$ component of 
angular momentum per unit mass is  $L = |r v_\theta|$ and computed from the a star's galactocentric
radius  $r$ and tangential velocity $v_\theta$.  
We compute the energy difference
approximating $E_c(L)$ for a flat rotation curve with circular velocity $v_c = 230 $ \kms.
We use $V_{circ,\odot}$ specifically for the LSR's circular velocity and
$v_c = V_{circ,\odot}$ for the circular velocity when we approximate
the Galaxy with a flat rotation curve.

Figure \ref{fig:LE} shows the number of stars in different angular momentum and energy
difference bins.  This figure is similar to a Toomre diagram where $u^2 + w^2$
is plotted against  $v$ (see Figure 2 by \citealt{bensby07}).
Our histograms in Figure \ref{fig:LE} 
show the number of stars in bins that are 5 \kmskpc wide in angular momentum
and 0.001 wide in the energy difference (approximately the orbital eccentricity squared or  $e^2$).
Stars are restricted to 
 distances $d< 1 $ kpc and  latitudes $|b| < 45^\circ$ 
 (and the GALAH survey itself contains few stars with $|b|<10^\circ$).
There is no cut in galactic longitude.  
The y-axes of each panel show the angular momentum in units of \kmskpcns.
The x-axes are unitless as the energy difference has been normalized by $v_c^2$.
The number of stars used to make each histogram in Figure \ref{fig:LE}
is shown in green on the lower left of each panel.
Again the histograms are normalized so that they integrate to 1.
Three normalized histograms are shown, the top one is for high metallicity stars with 
[Fe/H] $> 0.2$, the middle one is for lower metallicity stars with  [Fe/H] $<-0.1$
and the bottom one is for high $\alpha$-element stars with [$\alpha$/Fe] $> 0.2$. 
In the topmost histogram the right axis
shows the guiding radius $r_g$ computed from the angular momentum and assuming
a flat rotation curve ($r_g(L) = L/v_c $).  
The top axis shows the orbital eccentricity assuming
that the energy difference is  the square of the eccentricity. 

Figure \ref{fig:LE} shows that the high metallicity and $\alpha$-rich stars both
show extensions to low angular momentum, corresponding to stars coming
from the inner Galaxy.  At low metallicity there are more stars 
with higher angular momentum, corresponding to stars coming into the solar
neighbourhood from the outer Galaxy.
The top panel lets us estimate the angular momentum
of a gap separating the Hercules stream from more circular orbits at about
$L_{gap} \sim 1600 $ \kmskpcns.
We now use the gap angular momentum value to estimate the location of the gap in the different local 
neighbourhoods that are seen at each galactic longitude in Figure \ref{fig:line}a.

\subsection{The Hercules stream seen in different directions}
\label{sec:herlon}

\begin{figure*}
\includegraphics[width=18.5cm,trim={5mm 0 0mm 0},clip]{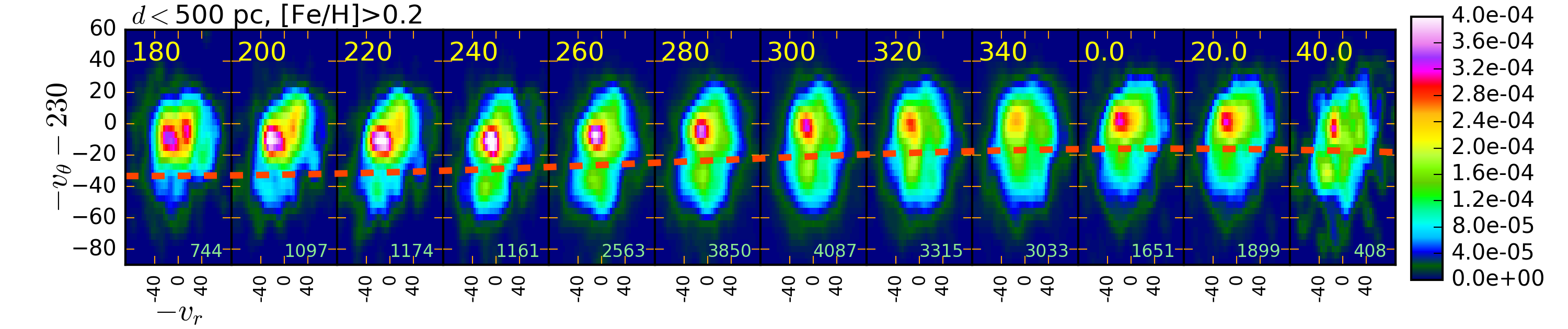}
\caption{The $uv$ velocity distributions of stars with distances  $d<500$pc and
 [Fe/H] $ > 0.2$ as  a function of galactic longitude.  The orange line (computed
 with equation \ref{eqn:vl})
 corresponds to a gap angular momentum value of $L_{gap} = 1640$ \kmskpc. 
The line curves because the galactocentric radius of each neighborhood depends  
on the viewed longitude.
%b) The bottom panel shows the $vw$ velocity distribution.
%On each figure we also plot
%a line showing a constant angular momentum value of 1640 \kmskpc and computed
%with equation \ref{eqn:vl}.
%The line was made assuming that the central position of each neighbourhood is 340 pc away
%and a rotation velocity of 230 \kms.
The line suggests that the gap between the Hercules stream and circular orbits
is set by a particular angular momentum value, as would be expected with a bar resonant model.
\label{fig:fn_line}}
\end{figure*}

\begin{figure}
\includegraphics[width=8.5cm,trim={5mm 0 0mm 0},clip]{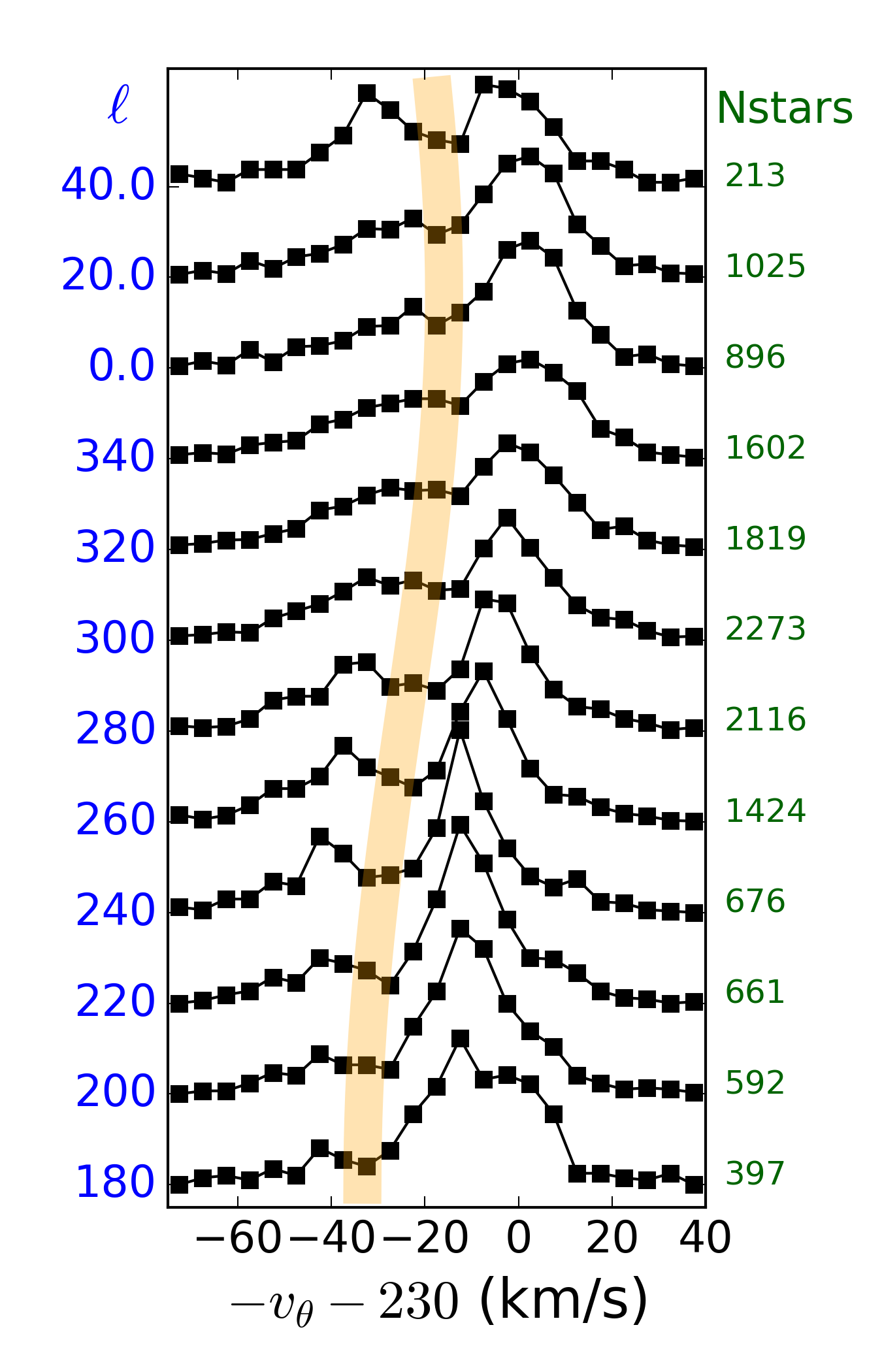}
\caption{Normalized tangential velocity distributions are shown for stars at each galactic
longitude given in blue on the left axis.  The numbers of stars in each histogram
is shown on the right axis.  The vertical range is not shown on either axis.   
These distributions
are similar to that shown in Figure \ref{fig:vhist} except they are
for stars at different galactic longitudes.
The thick orange line is computed
with equation \ref{eqn:vl} and
corresponds to a gap angular momentum value of $L_{gap} = 1640$ \kmskpcns. 
The Hercules stream forms the peaks at around $v\sim -40$ \kms
that are prominent near galactic longitude $\ell = 270^\circ$.
\label{fig:vhist_line}}
\end{figure}

Each velocity distribution shown in Figure \ref{fig:line}a was constructed of
stars from a different region in galactic longitude.  The $x,y$ position in galactic coordinates
of the centre of each region can be estimated from the central galactic longitude $l$
and the mean distance of the stars from the Sun, $d_n$;
\begin{align}
 x(l) &= d_n \cos l  - R_\odot \nonumber \\
 y(l)  &= d_n \sin l.
\end{align}
The galactocentric radius of the neighbourhood is estimated with $r(l) = \sqrt{x(l)^2 + y(l)^2}.$
The location of the gap in $v$ is set by its angular momentum with gap velocity as a function
of galactic longitude
\begin{equation}
v_{gap} (l) = \frac{ L_{gap}}{  r(l)} - v_c, \label{eqn:vl}
\end{equation}
where we use a circular velocity of $v_c = 230$ \kms.
Figure \ref{fig:fn_line} shows  the curve $v_{gap}(l)$  plotted on top of 
the ($u,v$) histograms we showed previously in Figure \ref{fig:line}a
for the high metallicity stars with distances $d<500$ pc.  
The curve was computed using a distance of $d_n = 340$ pc
which is the median distance from the Sun of stars used to make the histograms.
%Figure \ref{fig:fn_line}b  shows $vw$ velocity distributions
%generated in the same way as the $uv$ velocity distributions but using the 
%vertical velocity component $w  = v_z$ instead of the radial velocity component.
%The gaps are somewhat easier to see in the $vw$ velocity distributions
%than the $uv$ velocity distributions because the dispersion in $w$ is lower
%than that in $u$.
In Figure \ref{fig:vhist_line} we show normalized tangential velocity distributions 
(similar to that shown in Figure \ref{fig:vhist}) but constructed for stars
at different galactic longitudes.  Each distribution is restricted to stars with [Fe/H]$>0.2$, $d<500$ pc,
$-40<u<10$ \kms,
and galactic longitude within $20^\circ$ of the value given in blue on the left axis.
The numbers of stars in each histogram are given in green on the right axis.
The orange line shows the gap $v$ value computed using equation \ref{eqn:vl}.

Figure \ref{fig:fn_line} and Figure \ref{fig:vhist_line} show that the gap separating the Hercules
stream from circular orbits is reasonably well approximated
using the angular momentum value of $L_{gap} = 1640 \pm 40$  \kmskpc 
where we have estimated the error by eye after plotting curves for higher and lower
angular momentum values.  

Is the angular momentum value of the gap  $L_{gap} = 1640 \pm 40$  \kmskpc 
consistent with fast and slow bar models for the Hercules stream?
For the fast bar, the Hercules stream is caused by the outer Lindblad resonance  
\citep{kalnajs91,dehnen99b,dehnen00}
which is important for stars with angular rotation rate near  
$\Omega$ that satisfies a resonant condition
\begin{equation}
2(\Omega - \Omega_b) + \kappa \sim 0.
\end{equation}
For a flat rotation curve with $\kappa/\Omega = \sqrt{2}$ the resonant condition is equivalent to 
\begin{equation}
\Omega_b/\Omega = 1 + \frac{1}{\sqrt{2}} \approx 1.71.
\end{equation}
Multiplying by $v_c^2$ and setting $L_{gap} = v_c^2/\Omega$ for the resonant orbits with
angular rotation rate $\Omega$,  we can manipulate this expression to estimate
\begin{equation}
L_{gap} \approx \left(1 + \frac{1}{\sqrt{2}} \right) \frac{v_c^2}{\Omega_b} . \label{eqn:Lg}
\end{equation}
Using a fast bar pattern speed
$\Omega_b = 56 $ \kmsikpc \citep{antoja14} and circular velocity $v_c = 230$ \kmsns,
equation \ref{eqn:Lg}
gives a gap angular momentum of 1613 \kmskpcns,  that is consistent with
our estimate for the gap angular momentum value (1640; estimated
in Figure \ref{fig:LE} and shown as a line in Figures \ref{fig:fn_line} and \ref{fig:vhist_line}).
Our estimated gap angular momentum is consistent with previous estimates of $\Omega_b$ 
for a fast bar pattern speed and an outer Lindblad resonance model for the Hercules stream.

For the slow bar, with an estimated bar pattern speed of $\Omega_b = 39$ \kmsikpc \citep{perez17},
the Hercules stream is associated with stars at corotation and librating about a Lagrange point.
The angular momentum associated with corotating orbits
\begin{equation}
L_{CR} = \frac{ v_c^2}{\Omega_b}  = 1356\ {\rm km\ s}^{-1}\ {\rm kpc},
\end{equation}
which is lower than our estimated gap angular momentum of 1640 \kmskpcns.  
The Hercules stream stars themselves
have somewhat lower angular momentum values of about 1550 \kmskpc than our estimated gap value,
and both of these exceed the angular momentum associated with corotation
in the slow bar model.
Our estimated gap angular momentum does not support a corotation model
for the Hercules stream. Perhaps more careful orbital modeling is needed to
account for the velocity distribution as the orbits
in the corotation region have high amplitudes of libration
about the Lagrange points (see \citealt{perez17,portail17a,portail17b}).

The Hercules stream was detected 
using APOGEE giant stars that are a few kpc away and seen along a galactic
longitude of $270^\circ$ \citep{hunt18}.
%in APOGEE giants may not be consistent
%with the dependence on galactocentric position seen in the GALAH data. 
Stars that are 3 kpc in the 270 degree direction would have a galactocentric
radius of 8.5 kpc, outside that of the Sun.  At this
radius, equation \ref{eqn:vl} gives a $v$ value of -40 \kms 
for the gap velocity, somewhat larger than the gap $v \approx -27$ \kms at $\ell=270^\circ$
estimated for the nearer GALAH stars. 
We made a similar set of histograms for high metallicity GALAH stars with distance between 1 and 3 kpc. 
In those the Hercules stream is less noticeable and the gap separating
 it from the stars on circular orbits not visible
in the histograms.  However,
 there are only 1500 stars in the histogram at $\ell = 260^\circ$ as  the magnitude limited GALAH survey
contains fewer giants than main sequence stars, so the lack of structure is not significant.
Stars 3 kpc away from the Sun in the $270^\circ$ direction are located 
at a galactic azimuthal angle $\theta$
of $20^\circ$ (measured from the Sun galactic centre line) and the azimuthal angle
in the  Galaxy  too should affect the velocity distribution.
We do not confirm, but neither do we dispute the detection of the Hercules stream
in giant stars a few kpc from the Sun by \citet{hunt18}.

\begin{figure*}
\includegraphics[width=18.5cm,trim={5mm 0 0mm 0},clip]{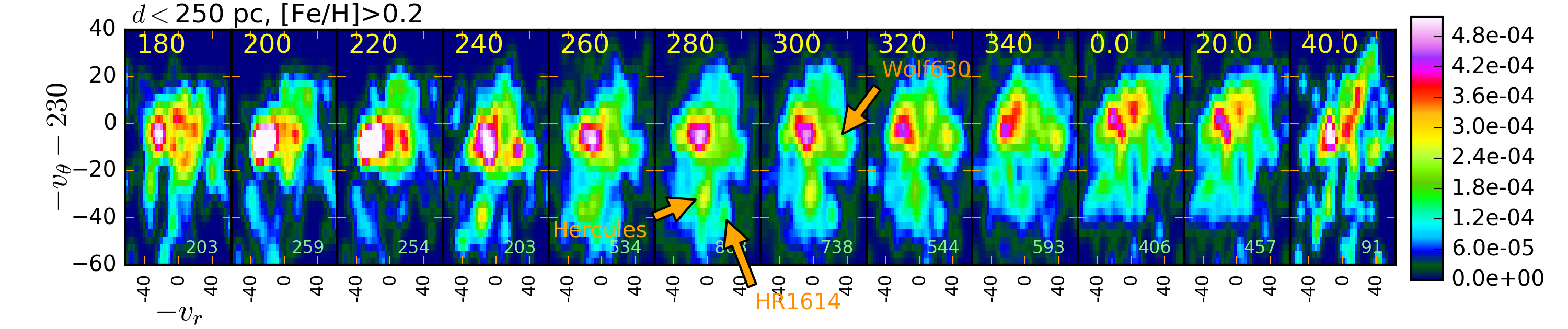}
\includegraphics[width=18.5cm,trim={5mm 0 0mm 0},clip]{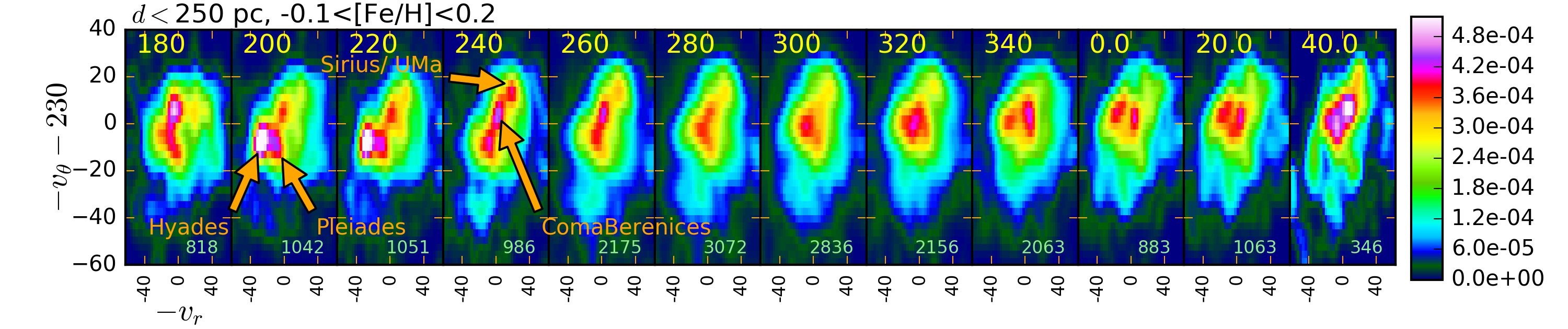}
\includegraphics[width=18.5cm,trim={5mm 0 0mm 0},clip]{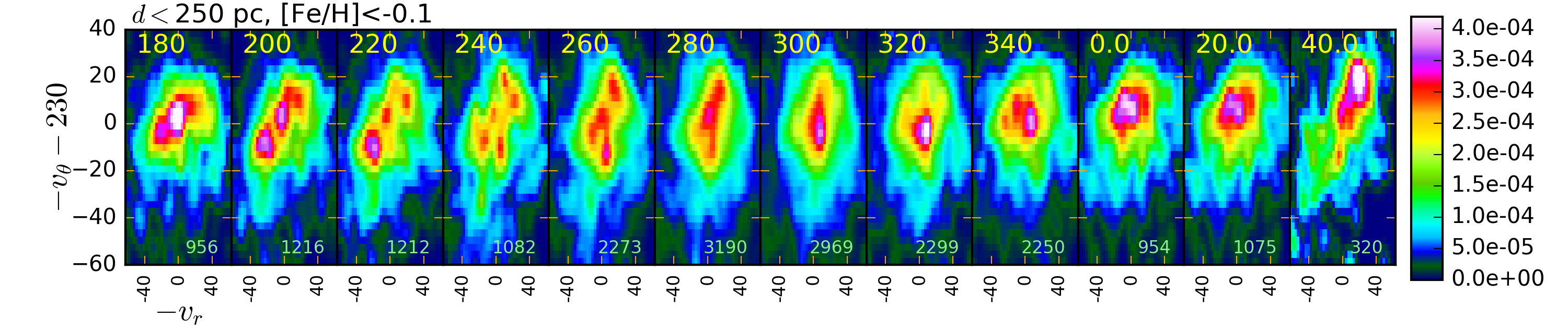}
\caption{Velocity distributions in $uv$ similar to those
in Figure \ref{fig:line} but for nearer stars  with distances  $d<250$pc.
a) For metal rich stars with [Fe/H] $ > 0.2$. 
b) For near solar metallicity stars $-0.1<$ [Fe/H] $ < 0.2$. 
c) For metal poor stars with [Fe/H] $ < -0.1$.
\label{fig:nn_line}}
\end{figure*}

\section{Structure in velocity distributions of nearer stars ($d<250$ pc)}
\label{sec:near}

The velocity distributions in Figure \ref{fig:line} do not obviously exhibit the 
clumps identified in the {\it Hipparcos} satellite observations of nearby and brighter stars
that corresponded to moving groups or stellar kinematic associations previously identified by 
Olin Eggen \citep{eggen58a,eggen58b,eggen58c,eggen65a,eggen65b,eggen65c,eggen95,eggen96} 
(see Figure 3 and Table 2 by  \citealt{dehnen98} and Figure 7 and Table 1 by \citealt{famaey05}).
To investigate this issue we look at velocity distributions for a nearer GALAH sample of stars.
Figure \ref{fig:nn_line} shows three sets of velocity distributions,
similar to those in Figure \ref{fig:line} but for distances $d<250$ pc instead of 500 pc.
Figure \ref{fig:nn_line}a (top row) shows metal rich stars with [Fe/H] $ > 0.2$, 
Figure \ref{fig:nn_line}b (middle row) shows near solar metallicity stars with $-0.1<$ [Fe/H] $ < 0.2$
and  Figure \ref{fig:nn_line}c (bottom row) show lower metallicity stars with [Fe/H] $ < -0.1$.
These histograms are constructed with 3 \kms square bins in ($u,v$) velocity components.
We did not make a similar figure for high $\alpha$-element stars due to the lower numbers
of stars in the histograms (which were noisy).

Clumps in velocity distributions are often designated by the name of 
a star or cluster that has similar space motions or by a direction on the sky where members have been
discovered \citep{eggen65c}.
These clumps are called moving groups, streams, stellar associations, 
or superclusters \citep{eggen65c,dehnen98,famaey05}.
In the velocity distributions shown in Figure \ref{fig:nn_line}, 
the moving groups called Hyades, Pleiades, Coma Berenices, and
Sirius/UMa by \citet{dehnen98} are visible and are labelled in the near solar metallicity row of Figure \ref{fig:nn_line}b.  We have labeled the Hercules stream in the metallicity rich 
histograms of Figure \ref{fig:nn_line}a. There is an extension to the Hercules stream 
at $(u,v) \sim (20,-50)$ \kms
which we tentatively identify with the HR1614 moving group,
and this too is labelled on  Figure \ref{fig:nn_line}a.
\citet{famaey05} labelled the Pleaides and Hyades moving groups discussed by \citet{dehnen98} 
as a single association called the Hyades-Pleiades supercluster, 
the Coma Berenices group discussed by \citet{dehnen98} 
approximately corresponds to their Y group, and the
Sirius/UMa group to their Sirius supercluster (as listed in their Tables 1 and 2 and shown in their Figure 9).
The Hyades, Pleiades, Sirius, Hercules, and
HR1614, groups were also identified in LAMOST survey data by \citet{liang17}.
%\citet{famaey05} divided the Y group into two  clumps?
The $v$ velocity components of our groups are lower by about 20 \kms 
from those reported by \citet{dehnen98,famaey05,liang17} as
our adopted value for the peculiar motion of the Sun 
and the galactic circular velocity at $R_\odot$ differs from theirs
(\citealt{dehnen98} used $V_{circ,\odot}= 200$ \kms and $v_{\odot,pec} = 5.2$ \kms both somewhat 
lower than our adopted values of $V_{circ,\odot}=230$ and $v_{\odot,pec} =10$ \kmsns).

Figure \ref{fig:nn_line}a shows an
additional peak in the velocity distribution of high metallicity stars near $(u,v) = (26,-7)$ \kms and
for $220^\circ < \ell <340^\circ$.  We tentatively identify it as Wolf 630
as its location is approximately consistent with the group of the same name identified by
\citet{liang17}.   The group is denoted Group 6  by \citet{dehnen98}. 
The space motion measured  by  \citet{bubar10} for the Wolf 630 moving group 
(their $(u,v) \sim (25, -36)$ \kmsns) is 
equivalent to our $(u,v) \sim (25, -16)$ \kms with our velocity convention.  This is sufficiently
near our peak's (26,-7) \kms to be consistent. 
We primarily see the Wolf 630 group in the nearby higher metallicity stars, though \citet{bubar10}
found a fairly broad metallicity distribution centered at near solar values (see their Figure 4). 

In the nearby GALAH stars, we recover the moving groups and streams previously
seen in {\it Hipparcos} observations.  
However stars observed by {\it Hipparcos} are brighter than a V mag of 7
and are much 
brighter than the GALAH survey stars.  There should be  little overlap in the two samples.
The GALAH detection of the these streams and groups confirms the existence of
these structures   in the solar neighbourhood's velocity distribution.

Figure \ref{fig:nn_line} shows that 
the moving groups or stellar associations identified by Olin Eggen and seen in the {\it Hipparcos} stars
are primarily seen in near solar or lower metallicities and at galactic longitudes $200^\circ < \ell < 260^\circ$,
though the Hyades group also contains higher metallicity stars.
The Hercules stream and the HR1614 moving group
at negative $v$ are also seen, but primarily in the high metallicity stars
and near galactic longitude $\ell \sim 270^\circ$.  
The Wolf 630 moving group group, near $v \sim 0$, is also most visible in the high metallicity stars
but at $240^\circ< \ell < 340^\circ$.

The Hercules stream is the only stream that is prominent in the more distant sample
(comparing Figure \ref{fig:line}a for $d<500$ pc to  Figure \ref{fig:nn_line}a for $d<250$ pc).
The stream is evident in local velocity distributions 
for galactic longitudes $\ell<240^\circ$ and $\ell>340^\circ$ but not
identified as a peak % in the high metallicity velocity distribution 
outside this range.
There is some ambiguity on the $u$ value of the Hercules stream.
\citet{dehnen98} and \citet{famaey05} identify two groups at $v \sim -50 $ \kmsns.
One is at $u \sim -40$ \kms and denoted Group 8 by \citet{dehnen98}, denoted Group
He by \citet{famaey05}, and consistent that denoted Hercules I by \citet{liang17}.
The other group is at $u  \sim -20 $ \kms and  is denoted
Group 9 by \citet{dehnen98},  group HV by \citet{famaey05} and Hercules II by \citet{liang17}.
The studies by \citet{dehnen98,famaey05} were both using {\it Hipparcos} observations so we could
expect similar clump identifications.  Using LAMOST survey data
\citet{liang17} identified two additional and weaker clumps 
associated with the Hercules stream.
% \citet{famaey05} found $v_{\odot,pec}$ the same and I have not yet found v_c
%The Hercules stream and the HR1614 are primarily seen in the metal rich stars
%and near galactic longitude $\ell \sim 270^\circ$.

The Hyades stream has the same velocities as the open cluster M67, though we have been careful
to remove a K2-HERMES  field that serendipitously contained M67 stars
in the near solar metallicity histograms (see Table \ref{tab:cuts} for the list of figures affected
by this cut).
The GALAH survey contains
few stars that are identified members of open or globular clusters \citep{kos18}.
The paucity of serendipitously observed cluster stars in the GALAH sample  is due
to logistics associated with fibre positioning and scheduling. 
So we are confident that the peaks seen in the 
nearby stars are not due to contamination by bound cluster members.  
This contrasts with the {\it Hipparcos}
survey which specifically targeted young clusters, including the Pleiades 
and Hyades clusters \citep{perryman97}.

In our velocity histograms or distributions we identify the Hyades, Pleaides, Coma Berenices,
Sirius/UMa, Wolf 630 and HR1614 streams or moving groups in the very nearby stars ($d<250$) and
these groups were much more difficult to see in the more distant $d<500$ pc sample.  
As the GALAH survey (neglecting targeted clusters) has
few cluster members, we can regard these moving groups as fine structure in the local velocity
distribution that is both metallicity dependent and dependent on position in the Galaxy, varying
over distances of a few hundred pc.  We infer that this structure varies on short distances
because the appearance of the velocity distributions varies as  a function of
 galactic longitude.

In the more distant samples (Figure \ref{fig:line}),
the velocity distributions appear smoother and with less fine structure than in Figure \ref{fig:nn_line}.
Errors in distances and tangential velocities, which are larger for the more
distant samples, could smooth out substructure in the velocity distribution.
Our more distant sample covers a wider volume 
causing coarse graining or averaging in the velocity distribution, and this too would blur substructure.
More distant stars tend to be located at higher elevations from the Galactic plane.  
The galactic longitude range of our stars $10^\circ < |b| < 45^\circ$  corresponds to
vertical distances  $87 \la |z| \la 350$ pc for stars at a distance $d=500$ pc
and $43 \la |z| \la 180$ pc for stars at $d=250$ pc.  At the larger distances of Figure \ref{fig:line}
the high galactic latitude directions should contain few thin disc stars.    
Stars more distant from the Galactic plane
may be less sensitive to perturbations associated with spiral structure, so velocity distributions at high $|z|$ 
may be smoother than those at low $|z|$.
Future GAIA observations will allow us to confirm our findings and search for similar fine substructure in 
more distant neighbourhoods.

%High metallicity stars:  the thing at $(u,v)$ = (70,-20) is mainly nearby and near anticentre 
%$180< \ell <210$ and above $\ell>0^\circ$. Could stretch between 35 and 180 where we are not plotting.
%Wider spread in radial velocities are primarily seen near anticentre.

\subsection{Peaks in the velocity distributions}
\label{subsec:peaks}

\begin{figure*}
\includegraphics[width=18.5cm,trim={5mm 0 0mm 0},clip]{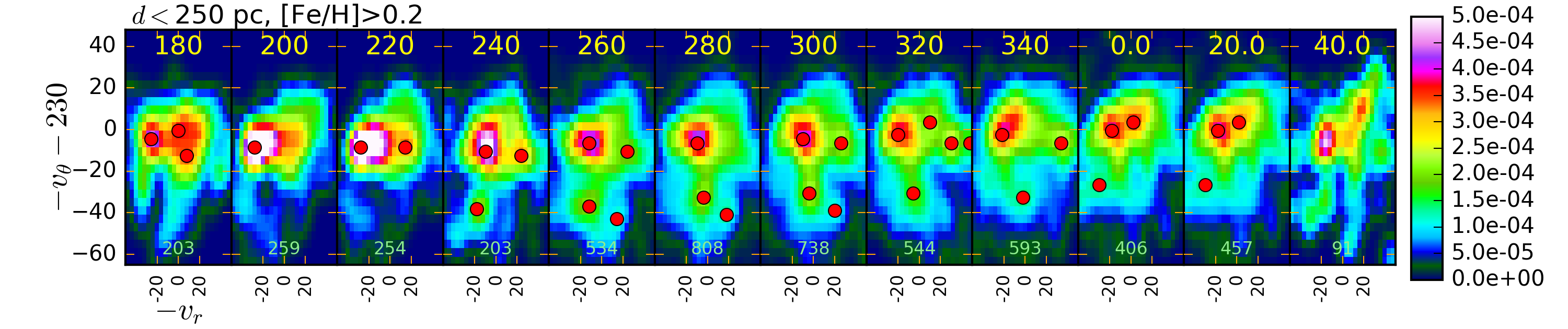}
\includegraphics[width=18.5cm,trim={5mm 0 0mm 0},clip]{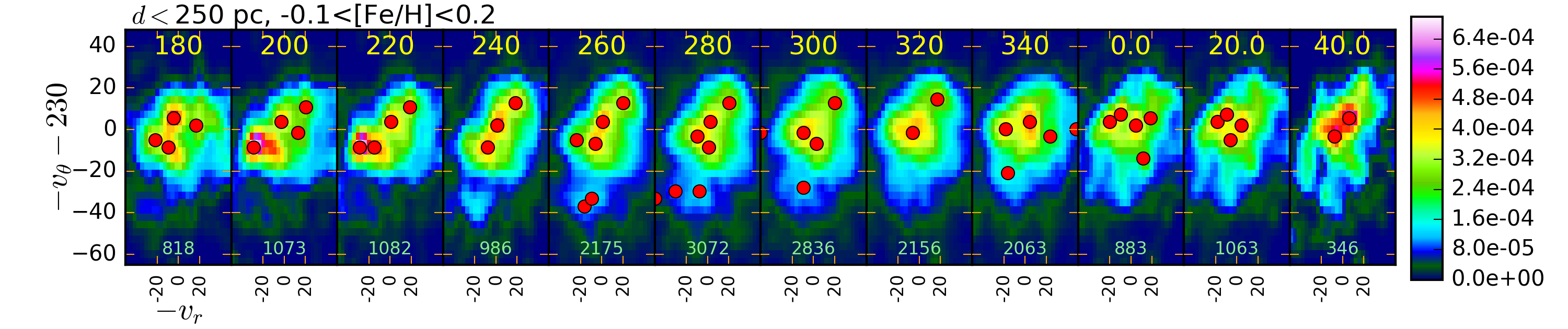}
\includegraphics[width=18.5cm,trim={5mm 0 0mm 0},clip]{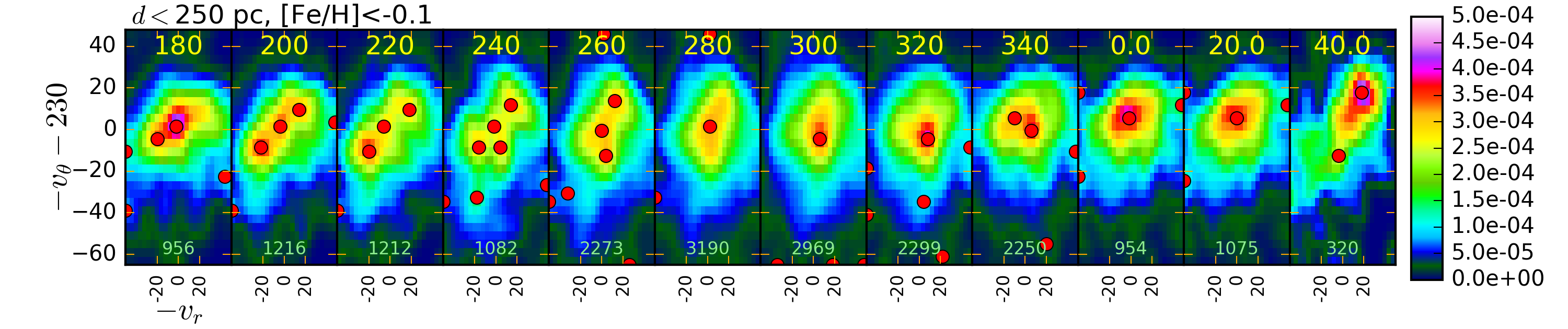}
\caption{Peaks are found in the $uv$ velocity distributions of nearby stars that are 
shown in Figure \ref{fig:nn_line}. 
Red dots show the ($u,v$) velocities of the peaks and they are plotted on top
of the velocity distributions.
\label{fig:nn_peak}}
\end{figure*}

To investigate how a clump in the local velocity distributions varies with location in the Galaxy, 
we  identify the velocities of peaks in the histograms.
We  smooth the histograms in Figure \ref{fig:nn_line} 
 by a few km/s.  We flag a peak in the velocity distribution if
 all adjacent bins are lower than that in the bin and if
the number of counts in the bin is higher than a threshold value that we adjusted
separately for each metallicity sample.  
The results of this unsophisticated peak finding routine are shown
in Figure \ref{fig:nn_peak} where we plot as red dots the position of peaks on top of the velocity distributions.  
The peaks are listed in Tables \ref{tab:tab1} -- \ref{tab:tab3} in Appendix \ref{sec:ap_peaks}.

The peaks shown in 
Figure \ref{fig:nn_peak} illustrate that the moving groups 
that are prominent in the velocity distributions are dependent on both viewing direction and
 metallicity. 
The Sirius/UMa group is primarily seen in solar or low metallicity stars,
confirming previous metallicity studies \citep{king05,tabernero17}, and is seen 
at galactic longitudes $200^\circ<\ell<300^\circ$.
The Coma Berenices group is primarily seen in low metallicity stars  in a similar galactic longitude range.
The Hyades group or supercluster is seen at all metallicities, confirming previous studies
\citep{tabernero12, pompeia11,desilva11}.
The Hercules stream is most prominent at higher metallicity
and near $\ell \sim 270^\circ$.    The stream is seen at most galactic longitudes
in the velocity distributions, but in many directions  it is a plateau in the velocity
distribution rather than a peak.

The peak finding routine finds an additional peak near the Hercules
stream but at lower $v$  and at positive $u$.  This clump could be 
 the HR1614 moving group \citep{eggen98,desilva07},
though the clump we see is near $u \sim +20 $ \kms  whereas previous studies 
find that the group is a narrow but tilted feature
in the uv plane that ranges from $u \sim -50 $  to $u \sim 20$ \kms with $v$ increasing 
from -70 to -50 \kms (see Figure 5 by \citealt{desilva07}). 
%Though \citet{eggen98,desilva07} 
This clump in our velocity distribution appears compact, suggesting
that its progenitor is an old and recently disrupted open cluster. If so its survival 
prior to disruption is a puzzle. Its age is estimated at 2 Gyrs \citep{desilva07} but even
a massive star cluster born in the disc is unlikely to survive that long \citep{fujii16}.
Streams or clumps in the velocity distribution are also seen in the low $v$ high $\alpha$ stars
at a similar $u$ value as the HR1614 clump.   If the high $\alpha$ stars (see Figure \ref{fig:line}d) 
contain similar structure to the high metallicity stars
then a bar induced mechanism for the HR1614 clump would be more likely than
a chemically homogeneous recently disrupted cluster.
As the GALAH survey continues, improved abundance measurements
(probing for chemical homogeneity) and observation of additional high $\alpha$ stars improving
the appearance of the velocity distribution,    would allow us to tell
these two possibilities apart.  
 
\begin{figure*}
\includegraphics[width=20cm,trim={15mm 0 0mm 0},clip]{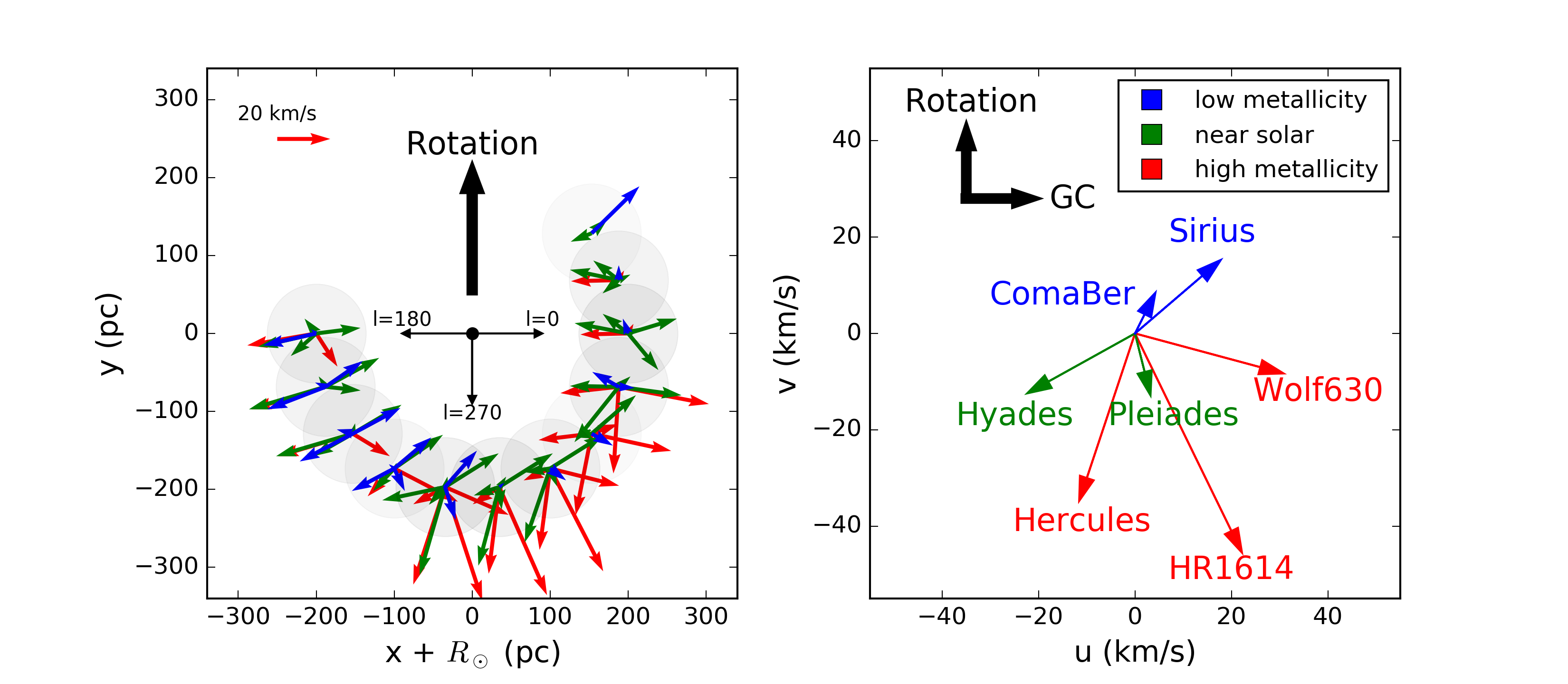}
\caption{a) Left panel.   Peaks seen in the $d<250$ pc $uv$ velocity distributions of Figures \ref{fig:nn_line}
and \ref{fig:nn_peak} are drawn as a function of 
position in the Galaxy.   The $x,y$ axis coordinates are centered about the Sun with positive
$x$ toward the Galactic centre and positive $y$ in the direction of rotation.
Green arrows show the velocity vectors for peaks in the velocity distribution 
for solar metallicity stars.   Blue arrows are for peaks in the lower metallicity set ([Fe/H] $<-0.1$)
and red arrows are for peaks in the higher metallicity set  ([Fe/H] $> 0.2$).
Coloured arrows are shown for the peaks we identified in Figures \ref{fig:nn_peak}.
Arrow length depends on the $u,v$ velocity vector length.  
The red arrow on the upper left shows a velocity vector of length 20 \kmsns.
Coloured arrows pointing to the right
correspond to motions toward the Galactic centre and arrows pointing upward
correspond to motion faster than Galactic rotation.
Neighbourhoods are shown as grey filled circles at a distance of 200 pc from the Sun and at
galactic longitudes ranging from $\ell = 180^\circ$ (on the left) to $40^\circ$ on the upper right.
Each grey circle represent the location of stars that we see in a particular direction.
The smaller black arrows in the centre show the directions of the different galactic longitudes.
b) Right panel.  The illustration  shows the directions of velocity peaks 
and we have labelled them with the names of moving groups or streams.
On this panel thick black 
arrows show the direction of rotation and the direction of the Galactic centre.
\label{fig:arrow}}
\end{figure*}

\begin{figure*}
\includegraphics[width=20.0cm,trim={15mm 0 0mm 0},clip]{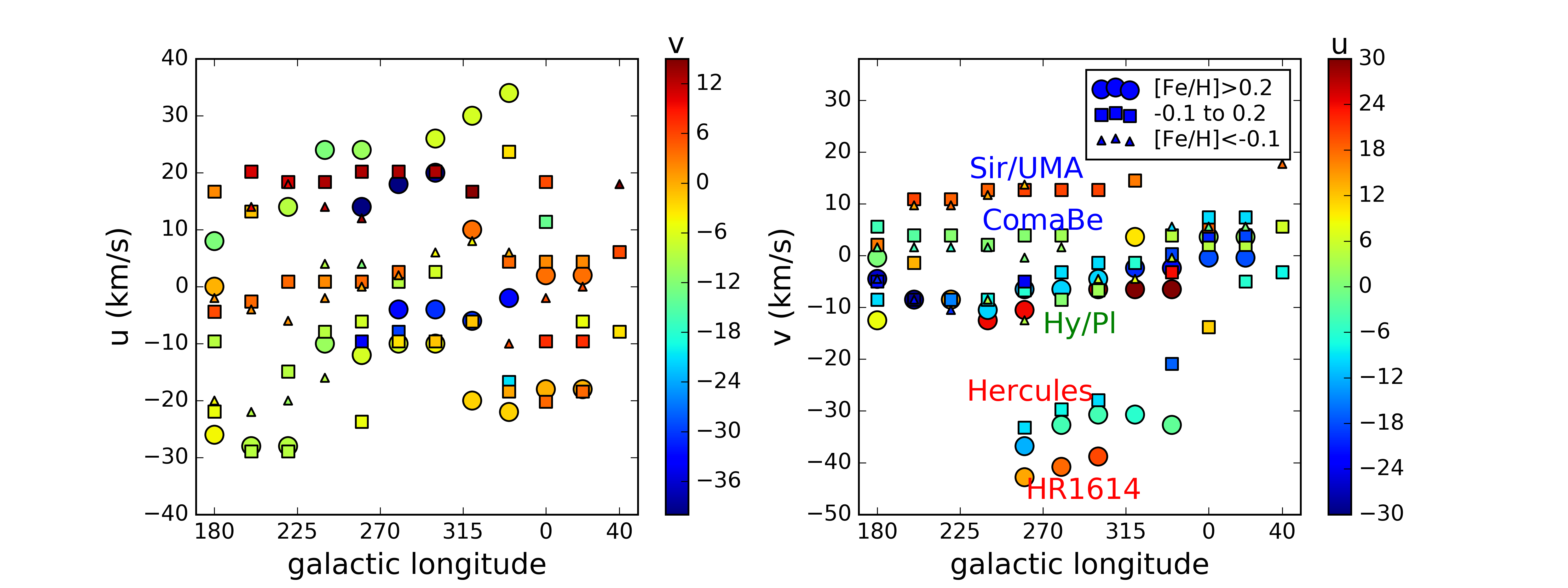}
\caption{Peaks seen in the $d>250$ pc $uv$ velocity distributions of 
Figures \ref{fig:nn_line}
and \ref{fig:nn_peak}
have $u = -v_r$ velocity component plotted
as a function of galactic longitude in the left panel. The colour of the points depends on the $v$ 
velocity (with $v = -v_\theta- 230$\kms as before) and with numbers on the colourbar in \kmsns.
In the right panel, the peak $v$ velocity is plotted as a function of galactic longitude and with colour
set by the $u$ velocity.  In both panels,
small triangles are for peaks in the lower metallicity set of histograms ([Fe/H] $<-0.1$),
squares are for near solar metallicities  ($-0.1 < $ [Fe/H] $<0.2$)
and large circles are for peaks in the higher metallicity set  ([Fe/H] $> 0.2$).
Sirius/UMa, Coma Berenices, Hyades/Pleiades, Hercules and HR1614
streams or moving groups are most easily identified in the panel on the right where
they are labelled. 
The peak $v$ velocities of each group varies as a function of galactic longitude, with $v$
increasing with increasing longitude over the range (in galactic longitude) 
where the group is distinctly seen 
in the local velocity distribution.
The Wolf 630 group can be seen on the right panel as red circles in the same region as Hyades and Pleiades.
The Hercules stream (with negative $u$) and the HR1614 group (with positive $u$) 
can be seen in the left panel as blue circles near galactic longitude $\ell=270^\circ$.
\label{fig:uv_lon}}
\end{figure*}

Each peak identified in the histograms in Figure \ref{fig:nn_line}
has specific $u,v$  values and these are a velocity
vector or a direction of motion.  To visualize the directions of motion we plot
the peaks as vectors in Figure \ref{fig:arrow}. 
%and their $u$ and $v$ components
%against galactic longitude in Figure \ref{fig:uv_lon}.  In Figure \ref{fig:arrow}
The $x,y$ axis coordinates are centered about the Sun with positive
$x$ toward the Galactic centre and positive $y$ in the direction of rotation.
The colour of the vectors is set by the metallicity with
red corresponding to peaks seen in higher metallicity stars, green for near solar metallicity stars
and blue for lower metallicity stars.
Coloured arrows in Figure \ref{fig:arrow} 
are shown for most of the identified peaks in Figures \ref{fig:nn_peak}  (we excluded
a few high velocity ones that were at positions with low numbers of stars in the histograms).
Coloured arrows pointing to the right
correspond to motions toward the Galactic centre (with positive $u$ or negative
$v_r$) and arrows pointing upward
correspond to motion faster than Galactic rotation (positive $v$).

In Figure \ref{fig:arrow}, 
neighbourhoods in the Galaxy are shown as grey filled circles at a distance of 200 pc from the Sun 
(just within  the 250 pc limit for the histograms of Figure \ref{fig:nn_line}).  
We show a grey circle  for each central
galactic longitude (for which we made histograms)
and these range from $l=180^\circ$ on the left to $l=40^\circ$ on the upper right.
The grey circles represent the neighbourhoods
and the coloured arrows represent the space motions
of stars in these neighbourhoods.
The velocities of the moving groups are shown as directions
in an illustration in the right panel of Figure \ref{fig:arrow}. 

Figure \ref{fig:arrow} shows smooth variations in the directions and lengths of 
space motions that are peaks in the local velocity distributions.  
For example Hyades and Sirius/UMa moving groups in the low metallicity
stars (blue arrows) are primarily seen on the left (at galactic longitudes  $180 < \ell < 280^\circ$)
but the direction of each arrow smoothly varies between $\ell = 180$ and $280^\circ$.
The Sirius/UMa group velocity increases in $v$ and decreases in $u$ with increasing longitude $\ell$.
The Hercules stream and HR1614 moving group  peaks increases 
in both $u$ and  in $v$ with increasing galactic longitude.
The Hyades supercluster, seen at all metallicities, increases in $v$ with increasing galactic longitude $\ell$.
This figure illustrates in a different way that structure in velocity distributions varies as a function
of location in the Galactic disc.

In Figure \ref{fig:uv_lon} $u$ and $v$ velocity components for the same peaks
seen in velocity distributions of
Figures \ref{fig:nn_line} and \ref{fig:nn_peak} are plotted against galactic longitude, $l$.
Small  triangles show peaks in the low metallicity stars,  squares show the near solar metallicity stars
and large circles show the higher metallicity stars.  
In the left panel, showing $u$ vs $l$, points are coloured by their $v$ value and the colour
scale shown with the accompanying colourbar.
In the right panel, showing $v$ vs $l$, points are coloured by their $u$ value.
The moving groups and streams
are easiest to identify in the right hand panel showing $v$ versus galactic longitude.
Over the range in galactic longitude where a peak is seen, peak $v$ velocities tend to
increase with increasing galactic longitude.   Trailing spiral arms might exhibit such a trend,
though a similar trend is seen in the Hercules stream and that one is associated with the Galactic bar.
The point colours make some streams stand out. 
The Wolf 630 group can be seen on the right panel as red circles in the same region as the
Hyades and Pleiades groups.
The Hercules stream and the HR1614 group can be seen in the left panel 
as blue circles near $l=270^\circ$.

The HR1614 group peaks (the three lowest red points in the right panel of Figure \ref{fig:uv_lon}) 
have $u$ increasing with $v$.  Fitting a line to these three points
in the form $v = a + b u$
(as did \citealt{desilva07} for HR1614 candidate stars in their Figure 5) 
we estimate a slope of $b\approx $2/3, exceeding the slope of $b=$0.18 measured by \citet{desilva07}.
\citet{desilva07} measured the slope (in velocity space) of a local distribution of bright and nearby stars
whereas we have measured a slope in the way peak velocity components vary with galactic longitude
for stars a few hundred pc away from the Sun.
The two measurements are not equivalent.  The slope measured
by \citet{desilva07} was probably not due to stellar velocities varying systematically with distance
from the Sun.

The local neighbourhoods at $l=180^\circ$ and $l=0^\circ$ are separated by about 400 pc.
The variations in $u,v$ values for the different stream peaks have differences  of size
$\Delta \sim 10$  \kmsns.  Dividing the velocity change by the distance between neighbourhoods
gives us  gradient in velocity.   The gradient  in the velocity components for the stream peaks has a 
size scale of order $\frac{dv}{dx} \sim 25$ \kmsikpc.

A worry is that a possible bias in the distance estimates could induce direction dependent trends 
in the velocity components that we have misinterpreted in terms of spatial gradients in the velocity 
distribution.
If distances are over-estimated then $u$ would be higher at $l=270^\circ$ than the correct value
and $u$ would be correct at $l=0,180^\circ$ as it depends only on the radial velocity and not the 
proper motion.
The $v$ component would be high at $l=0,180^\circ$
and correct at $l=270^\circ$.
The trends for the peaks we show in 
Figure \ref{fig:uv_lon} are inconsistent with over- or under-estimated distances.    
If a star has $v_{tan}= 30$ \kmsns, an error in the distance
of 30\% would be required to cause a 10 \kms error in the star's velocity $u$ velocity component. 
Thus extreme distance biases would be required to influence the velocities sufficiently to 
mimic the trends we see in Figure \ref{fig:uv_lon}.

%Improvements in distances
%and proper motion measurements with the forthcoming GAIA data releases will allow better
%measurement of local gradients in the velocity distributions and so can confirm or refute
%variations we see in the velocity distribution over distances of a few hundred pc.

The Sun lies just inside the Local Spiral Arm (hereafter the Local Arm), 
is about 2 kpc away from and inside the Perseus arm, 
and  is outside (again about about 2 kpc away) from the Sagittarius arm 
(see for example Figure 2 by \citealt{xu16}).
The Local Arm, containing the Orion stellar association and sometimes called the Orion spur,
is the nearest spiral arm to the Sun \citep{oort58,russeil03,vazquez08,xu13,houhan14,xu16}.
This arm is usually discussed as a  spur or arm segment that may arch between 
the Perseus and Carina spiral arms \citep{vazquez08,houhan14}, however parallax measurements
of young stars with bright molecular maser emission suggest that the Local Arm may be  separate
and stronger than previously inferred \citep{xu13,xu16}.
The Sun is interior to the Local Arm
 by about 500 pc with nearest associated stars seen in the Galactic anticentre
direction.  
The Sun is nearest the trailing sides of the Local and Perseus arms, but nearest
the leading side of the Sagittarius arm.

N-body simulations containing spiral arms 
exhibit radial velocities that generally point outward on the trailing side of spiral arms (and vice versa
for the leading side) and rotational or tangential 
velocities which are slower than the mean on the trailing side of spiral arms 
(and vice versa for the leading side; \citealt{grand14,grand15}).
This gives $u<0,v <0$ in a neighbourhood on the trailing side and $u>0, v >0$ on the leading side.
Variations in  peak ($u,v$) for a moving group with galactic longitude could be a function of proximity
to a spiral arm ridge.   The gradients of mean velocity components with
position in a simulated galaxy (for example see Figure 2 by \citealt{grand15}) seem large
enough to account for the smooth variations or gradient of the highest peak which is usually labeled as
 the  Hyades cluster.   Its negative $u$ and $v$ values are consistent with being on the trailing side
of a spiral arm.   This and the proximity of the Local Arm suggest that smooth
variations in $v$ as a function of galactic longitude in the Hyades stream are caused by the Local Arm. 
One can choose a set of neighbourhoods, similar to those covered by the GALAH survey, 
but in the simulated galaxy of Figure 2 by \citet{grand15} and with
central position lagging  a spiral arm.
It may be possible to  choose the central position with respect to the spiral structure so that the peaks 
in the local velocity distributions of the simulation exhibits the near sinusoidal 
variation in $v$ with galactic longitude that we see in the Hyades stream peak velocities.

Proximity to a single arm
might account for variations in peak ($u,v$) values or of the entire velocity distribution 
as a function of position but would
not account for the multiple peaks in the velocity distributions and how they vary with position in the Galaxy.
Fine structure seen in local velocity distributions that varies over distances as short
as a few hundred pc is a surprise.  
Strong spiral arms could cause more than one clump and deviations in the velocity distribution across
short distances if there are local resonances with a spiral pattern (e.g, \citealt{quillen05})
or if multiple patterns are present in the disk (e.g., \citealt{quillen11}).  
The illustrations in Figure 11 \citet{quillen11} show how stars at different guiding radii and perturbed
by different patterns can be present in a single neighbourhood, giving two velocity clumps
in a local stellar distribution.
However, previous studies \citep{quillen11,grand15}
did not expect so much substructure in the velocity distributions at lower
velocities and the large gradients. 
%the sizes of variations in peaks in the velocity distribution 
%over only a few hundred pc.  
Many transient spiral patterns, simultaneously in the disc,
might give numerous clumps in the velocity distribution \citep{desimone04} and gradients
in their locations.  A bar model that  takes into account
relaxation following bar growth and evolution
also might be consistent with variations in the velocity distribution over short
distances such as a few hundred pc \citep{minchev10}. 
Or perhaps there are short wavelength spiral density waves 
traveling through the disc, similar to those seen in
self-gravitating N-body shearing sheet simulations \citep{toomre91,quillen18}
and these might cause substructure in local velocity distributions.
Smaller scale spiral structures such as 
branches, spurs, feathers, and arm-segments, are seen
in external spiral galaxies \citep{sandage61,lynds70,elmegreen80}, and these
too may manifest as substructure in local velocity distributions if they are present 
in the Galactic disc near the Sun \citep{skuljan99b}.
Up to this point we have primarily discussed dynamical mechanisms associated
with perturbations to the Galactic disc.   Chemically homogeneous dissolved star clusters
could also contribute to the sub-structure in local velocity distributions.

%\citet{xu16} estimates a pitch angle of 10-11$^\circ$, similar to the Perseus and Sagittarius  spiral
%arms (see their figure 2).
%The local arm  is seen in Galactic quadrants 1,2 and 3 
%\citep{russeil03,vazquez08,houhan14} and along galactic longitudes about $ 80 <l < 270^\circ$.

\subsection{Dependence of Nearby Velocity Distributions on Galactic Hemisphere}
\label{subsec:hem}

\begin{figure*}
\includegraphics[width=18.5cm,trim={5mm 0 0mm 0},clip]{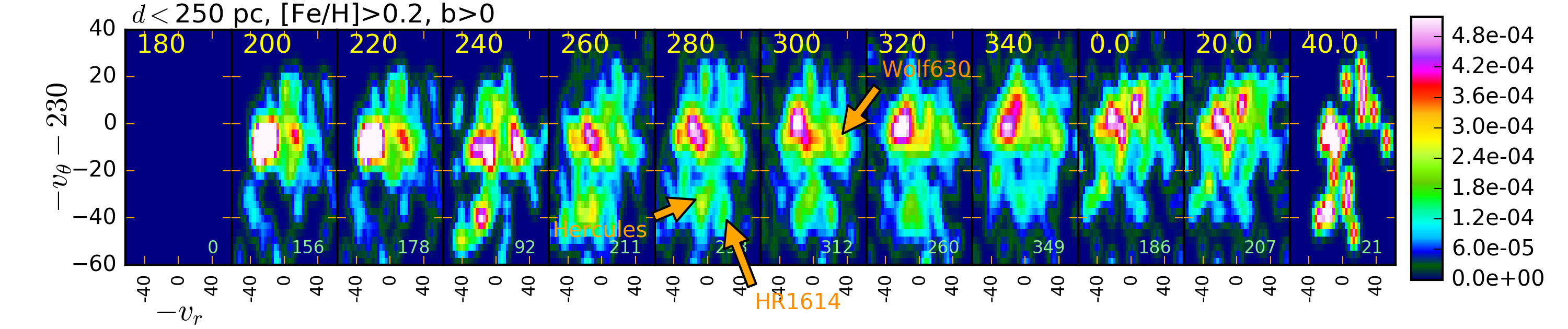}
\includegraphics[width=18.5cm,trim={5mm 0 0mm 0},clip]{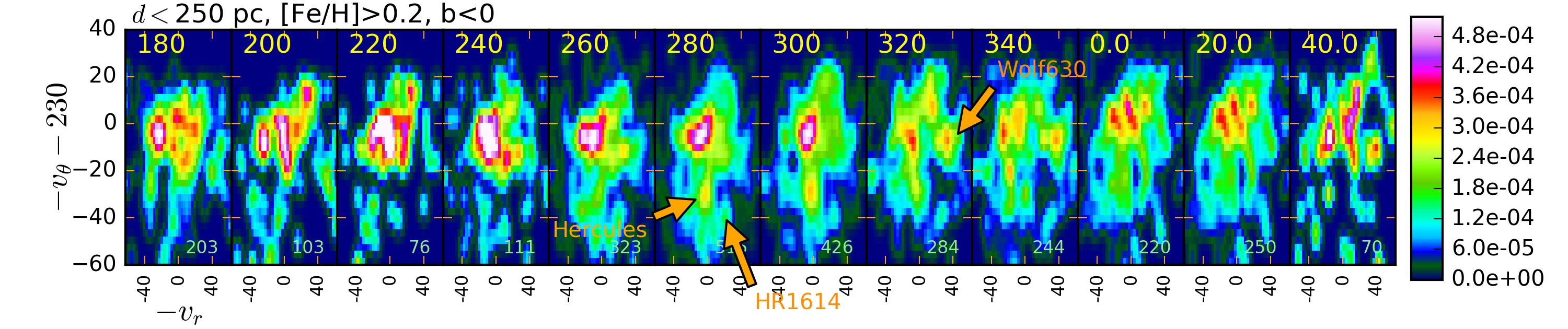}
\caption{Velocity distributions in $uv$ similar to those shown in Figure \ref{fig:nn_line}a for high metallicity stars,
but restricted to positive (top panel) and negative (bottom panel) galactic latitudes.
\label{fig:fnn_b}}
\end{figure*}

\begin{figure*}
\includegraphics[width=18.5cm,trim={5mm 0 0mm 0},clip]{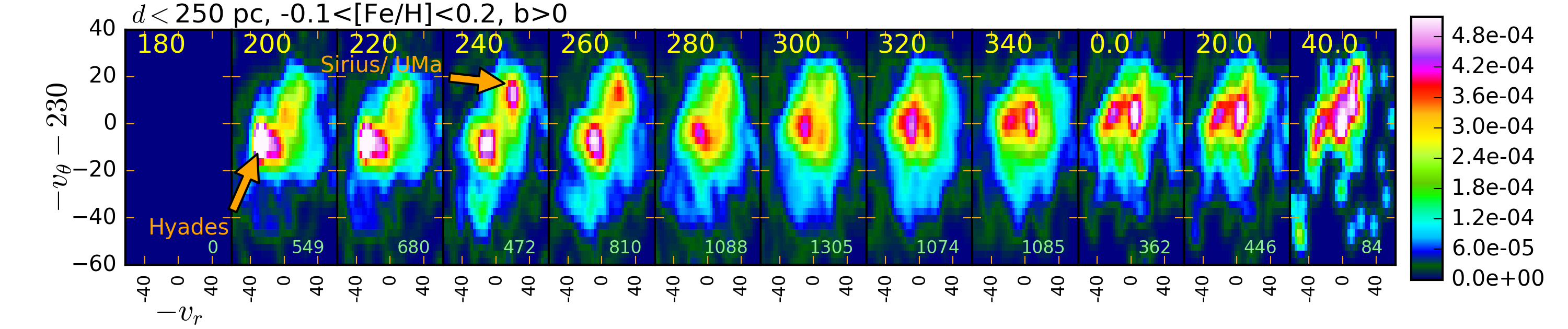}
\includegraphics[width=18.5cm,trim={5mm 0 0mm 0},clip]{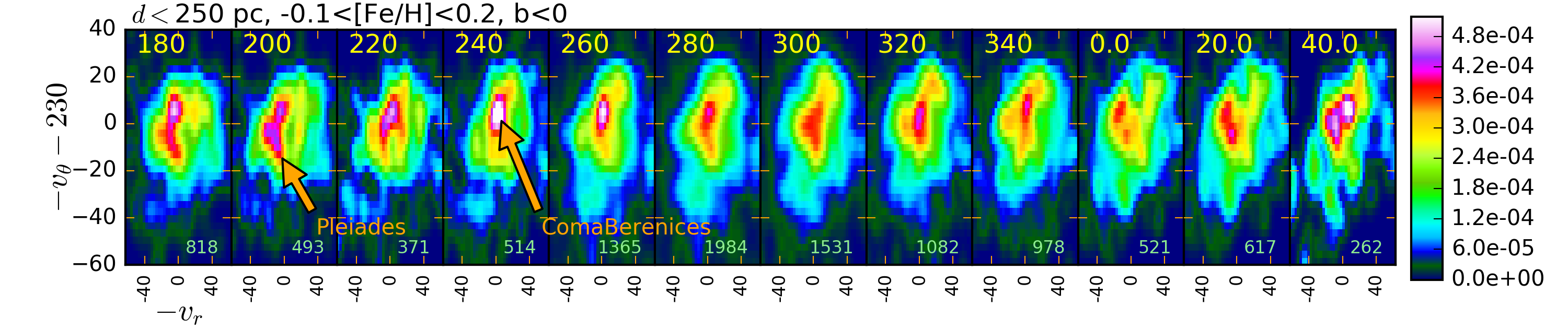}
\caption{Velocity distributions in $uv$ similar to those shown in Figure \ref{fig:nn_line}b
for near solar metallicity stars,
but restricted to positive (top panel) and negative (bottom panel) galactic latitudes.
\label{fig:mnn_b}}
\end{figure*}

\begin{figure*}
\includegraphics[width=18.5cm,trim={5mm 0 0mm 0},clip]{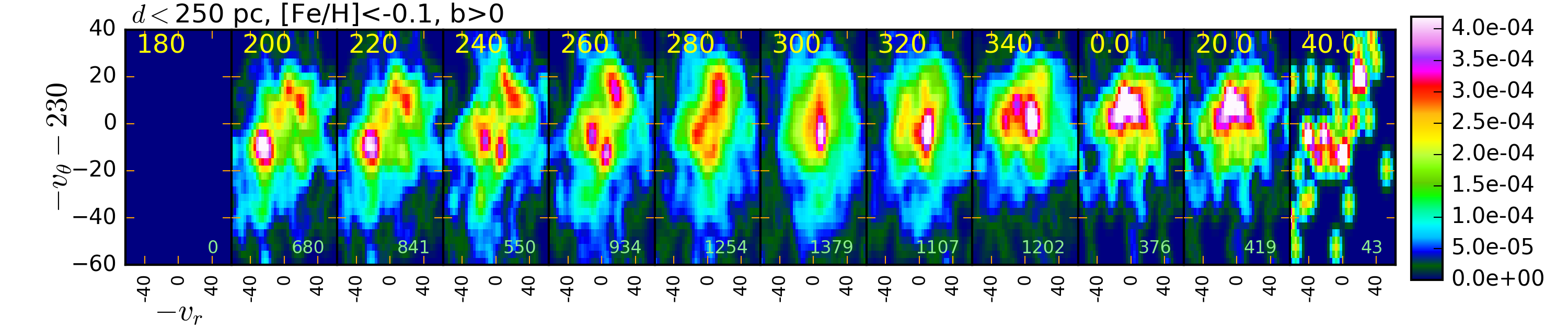}
\includegraphics[width=18.5cm,trim={5mm 0 0mm 0},clip]{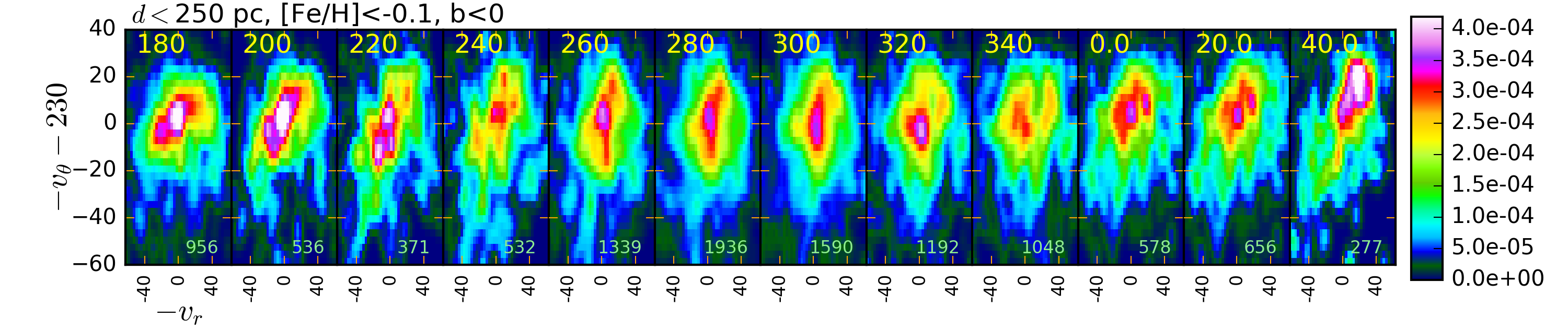}
\caption{Velocity distributions in $uv$ similar to those shown in Figure \ref{fig:nn_line}c for low metallicity stars,
but restricted to positive (top panel) and negative (bottom panel) Galactic latitudes.
The velocity distributions of low metallicity stars are the most sensitive to Galactic hemisphere,
with Coma Berenices moving group preferentially seen in the southern Galactic hemisphere
and Hyades and Sirius/UMa groups preferentially seen in the northern Galactic hemisphere.
\label{fig:lnn_b}}
\end{figure*}

Following a suggestion by Agris Kalnajs,
we recreated our velocity histograms for stars only at positive and only at
negative Galactic latitudes. 
The velocity distributions for $d<500$ pc  were not significantly sensitive to the Galactic hemisphere, so 
we again only show distributions for the $d<250$ pc stars.
Velocity distributions for both hemispheres in the high metallicity, near solar and lower metallicity
samples and for $d<250$ pc 
are shown in Figures \ref{fig:fnn_b}, \ref{fig:mnn_b}  and \ref{fig:lnn_b}.
We were surprised to find that
the Hyades and Sirus/UMa groups are predominantly seen at galactic latitudes $b>0^\circ$,
for $d<250 $ pc  near solar and lower metallicity stars.
The   
Coma Berenices group is predominantly seen at galactic latitudes $b<0^\circ$ and on the opposite side
as the Coma Berenices open cluster itself which is seen near the Galactic north pole.
We confirm the prior discovery by Agris Kalnajs (private communication) 
found previously using {\it Hipparcos} proper motions.

The dependence of peak positions in the velocity 
distribution on Galactic hemisphere is unexpected.  
Possible causes include small number statistics, as the GALAH sample is not uniformly distributed
across the sky.  There could
be variations in the distance distributions along different sight lines. If the peak velocities
depend on distance from the Sun,
then variations in the distance distributions
could cause variations in these peak velocities along different sight lines.
Alternatively, if there really is
small scale structure in the velocity distributions and strong gradients in the velocity
distributions within the Galaxy,
we would expect correlated stellar density variations.  In other words,
we should also see changes
in the number density of stars or stars per unit volume in the Galaxy.
There may also be correlated vertical velocity variations.
Correlated variations in stellar number density and vertical motions are predicted if
 correlated vertical and radial epicyclic motions in the disc are induced
by perturbations from dwarf galaxies in the outer Galaxy \citep{delavega15} 
or vertical breathing and bending waves travel through the disc \citep{widrow12,widrow14,widrow15}.

\begin{figure*}
\includegraphics[width=5.6cm,trim={10mm 0 0mm 0},clip]{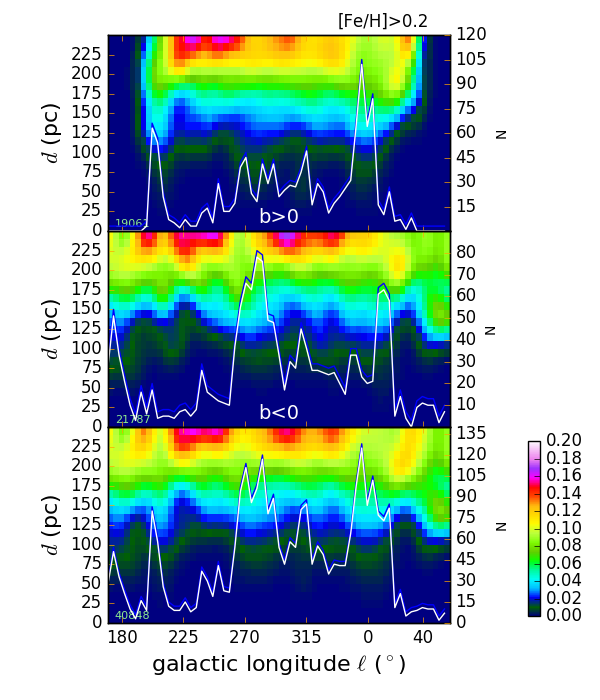}
\includegraphics[width=5.6cm,trim={10mm 0 0mm 0},clip]{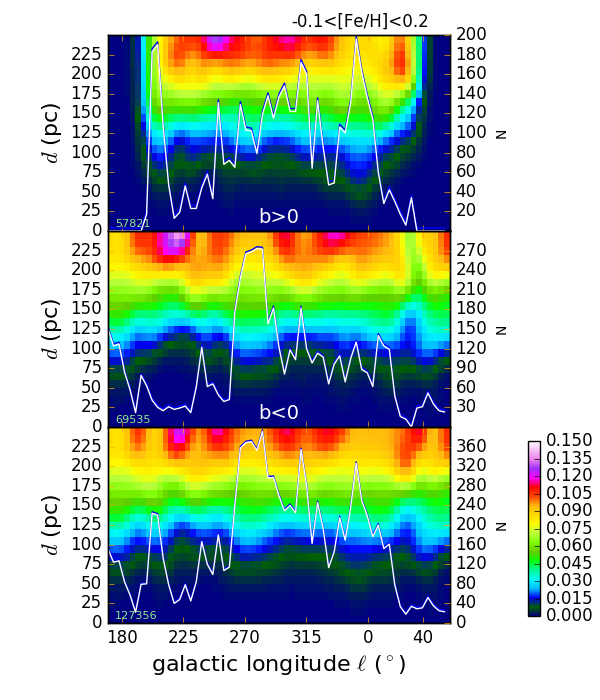}
\includegraphics[width=5.6cm,trim={10mm 0 0mm 0},clip]{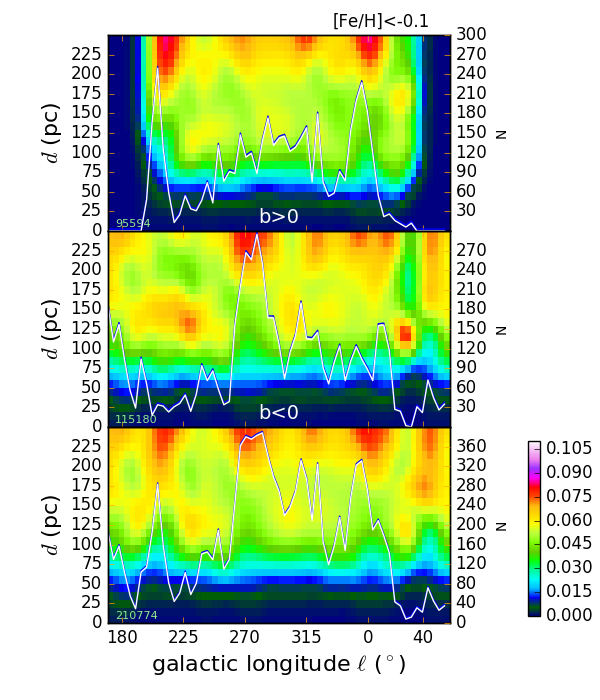}
\caption{Distance distributions of the GALAH sample as a function of galactic longitude and for the three 
different metallicity sets, with high metallicity on the left, near solar metallicity in the middle
and lower metallicity on the right.
Each panel shows a histogram showing numbers
of stars in distance and galactic longitude bins. 
The histograms have been
normalized so that each column, specified by its galactic longitude, sums to 1.  
The colourbars show
the fraction of stars at a given longitude per 10 pc wide bin in distance.  
Distance axis is on the left and the galactic longitude axis is on the bottom.
Galactic longitude bins are $4^\circ$ wide.
The total numbers of stars at each galactic longitude is plotted as a white line with y axis
showing the scale on the right.
Top panels refer to stars in the northern
Galactic hemisphere with Galactic latitude $b>0^\circ$, the middle panels for stars in the southern Galactic
hemisphere, and the bottom panels for stars in both hemispheres.
Only the low metallicity stars show large differences in the distance distributions between hemispheres.
 \label{fig:dgl}}
\end{figure*}

To investigate the effect of uneven sky coverage
and to see if there are sight-line dependent
differences in the distance distributions, we plot the distance distributions
for nearby GALAH stars in Figure \ref{fig:dgl}.
Our histograms count stars in bins that are 10 pc wide in distance  and $4^\circ$ wide in galactic longitude.
The axes for distance are shown on the left in pc.  The bottom axes
show galactic longitude $\ell$ in degrees.  The histograms have been
normalized so that each column, specified by its galactic longitude, sums to 1.
The normalization makes it easier to compare distance distributions at different longitudes.
The colourbars show
the fraction of stars (at each longitude) that is present in each 10 pc wide bin.  
The total numbers of stars with $d<250$ pc in each galactic longitude bin 
is plotted as a white and black line with y-axis
on the right showing the scale.
Top panels show stars in the northern
Galactic hemisphere (with Galactic latitude $b>0^\circ$), 
middle panels shows stars in the southern Galactic
hemisphere, and the bottom panels show stars in both hemispheres.
Three sets of histograms are shown. 
On the left we show the high metallicity stars with [Fe/H]$>$0.2. 
The middle
panels show near solar metallicity stars (-0.1$<$[Fe/H]$<$0.2) and the right panels show low metallicity stars
([Fe/H]$<$-0.1).   As before, only stars at latitudes $|b|<45^\circ$ and with
distance errors $\sigma_d/d < 0.5$ are used to construct the histograms.

The distance histograms in Figure \ref{fig:dgl} 
show that the low metallicity GALAH stars have the largest variations
in distance distribution between hemispheres.  
The region in galactic longitude with $200^\circ < \ell < 260^\circ$
has a ridge at $d\sim 130 $ pc in the southern Galactic hemisphere that
is absent in the northern Galactic hemisphere.
The number density variations are strongest
where there are fewest stars, which is a concern.  
%We could account for the differences in the velocity distributions seen in Figure \ref{fig:lnn_b}
%The Coma Berenices velocity vector is likely associated with  
Stars 130 pc away from the Sun 
and along $\ell \sim 230^\circ$ could tend to have velocity vector like the Coma Berenices group.
However the distance distribution in the near solar metallicity stars
do not show as large hemisphere sensitive variations as the 
low metallicity stars.  After removing the stars within 130 pc, we 
still see the Coma Berenices  stream in the southern hemisphere velocity
distribution in near solar metallicity stars, so this stream is not confined to very nearby stars.

In the previous section we inferred that there was evidence for variations in 
the velocity distributions (in $u,v$) over short distances. 
Variations or gradients in the velocity distributions, if they are real, probably would be
associated  with density variations (variations in the number of stars per unit volume)
over similar spatial scales.  It is tempting to interpret the differences in the distance
distributions in Figure \ref{fig:dgl} and in the lower metallicity stars
as evidence for variations in stellar number density.  However no similar density
variations are seen in the near solar metallicity 
stars and they too show the asymmetry in the velocity distributions
for $200^\circ < \ell < 260^\circ$.
%It is a concern that the variations in the distance histograms with hemisphere
%(and the associated changes in velocity peak locations)
%are largest where there are fewest stars.

At a distance of 200 pc, a star that has galactic latitude $b=10^\circ$ (the latitude limit
of the GALAH survey) is 34 pc above
the Galactic plane (where we have ignored the Sun's offset), and that at $b=45^\circ$
is 143 pc above the plane, but still within the thin disc scale height.
The differences in velocities seen between the northern and southern Galactic hemispheres
correspond to a vertical gradient in $v$ component velocity of about 10 \kms across 300 pc 
(approximately multiplying 134 by 2)
corresponding to 
about 30  \kmsikpcns.  This is 
 similar in size to the gradient in the plane we estimated for peaks in ($u,v$).
 
Does the position of the Sun, $25 \pm 5$ pc, (see \citealt{bland16} for a review of recent
measurements), 
above the plane in the Galactic north pole direction,
account for the differences in peak velocities seen in the two Galactic hemispheres?
Taking into account a 25 pc offset for the Sun, a star at 200 pc and at 
$b=10^\circ, -10^\circ, 45^\circ, -45^\circ$, 
would be at $z=9, -59, 118, -159$ pc, respectively.   These values lie well within the thin disc.
Strong vertical velocity gradients in the thin disc would still be required for the position of the Sun above
the Galactic plane to affect the viewed velocity distributions.  

\begin{figure*}
\includegraphics[width=18.5cm,trim={5mm 0 0mm 0},clip]{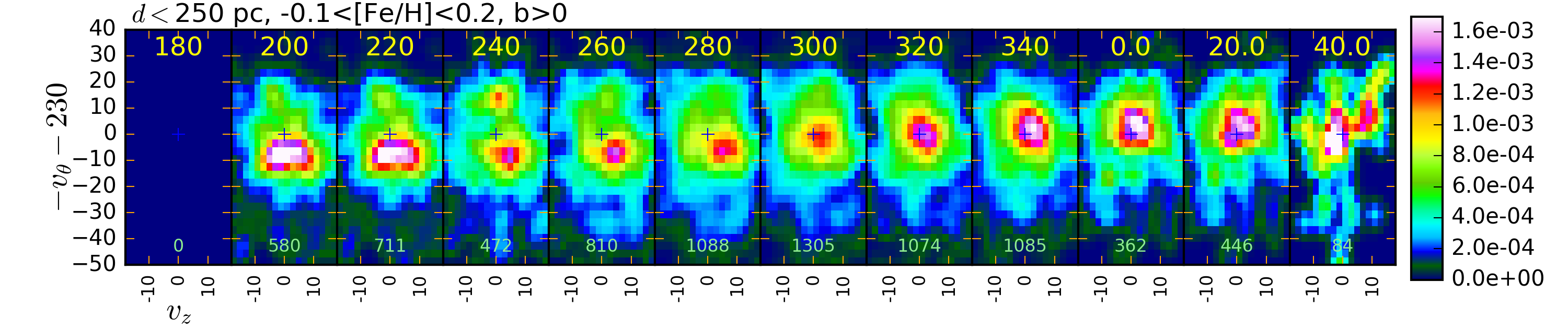}
\includegraphics[width=18.5cm,trim={5mm 0 0mm 0},clip]{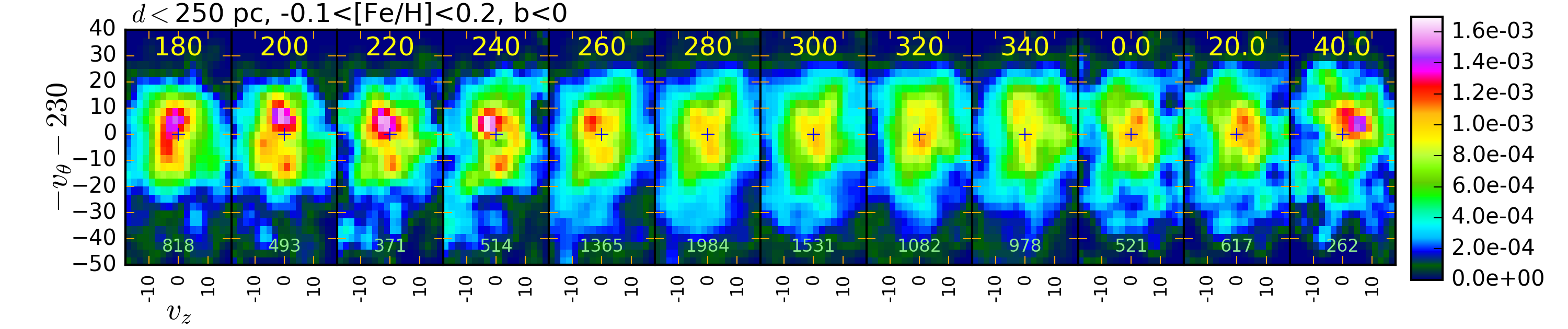}
\includegraphics[width=18.5cm,trim={5mm 0 0mm 0},clip]{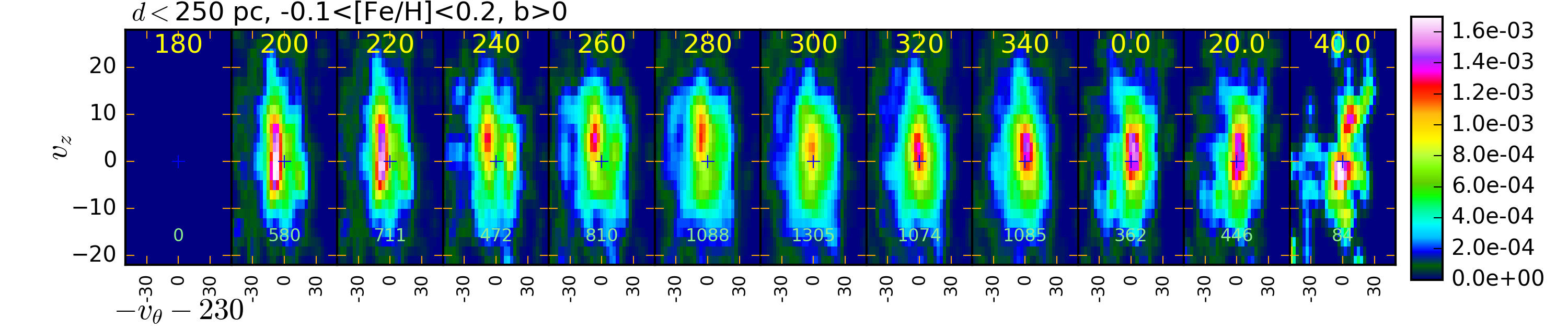}
\includegraphics[width=18.5cm,trim={5mm 0 0mm 0},clip]{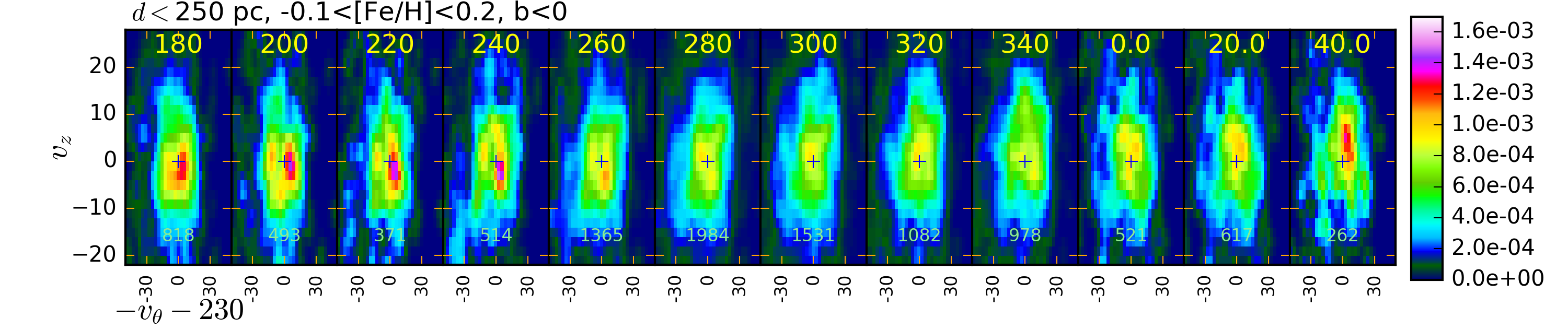}
\caption{Velocity distributions in $v,w$ for near solar metallicity stars.
The x-axes in the top and second from top panels are $v$ and y-axes are $w$.
The third from top and bottom panels are reversed with x-axes $w$ and y-axes  $v$.
The top and third from top panels show stars in the northern Galactic hemisphere (latitude $b>0$)
whereas the second from top and bottom panels show stars in the northern Galactic hemisphere.
 \label{fig:mmw_hem}}
\end{figure*}

\begin{figure*}
\includegraphics[width=20.0cm,trim={15mm 0 0mm 0},clip]{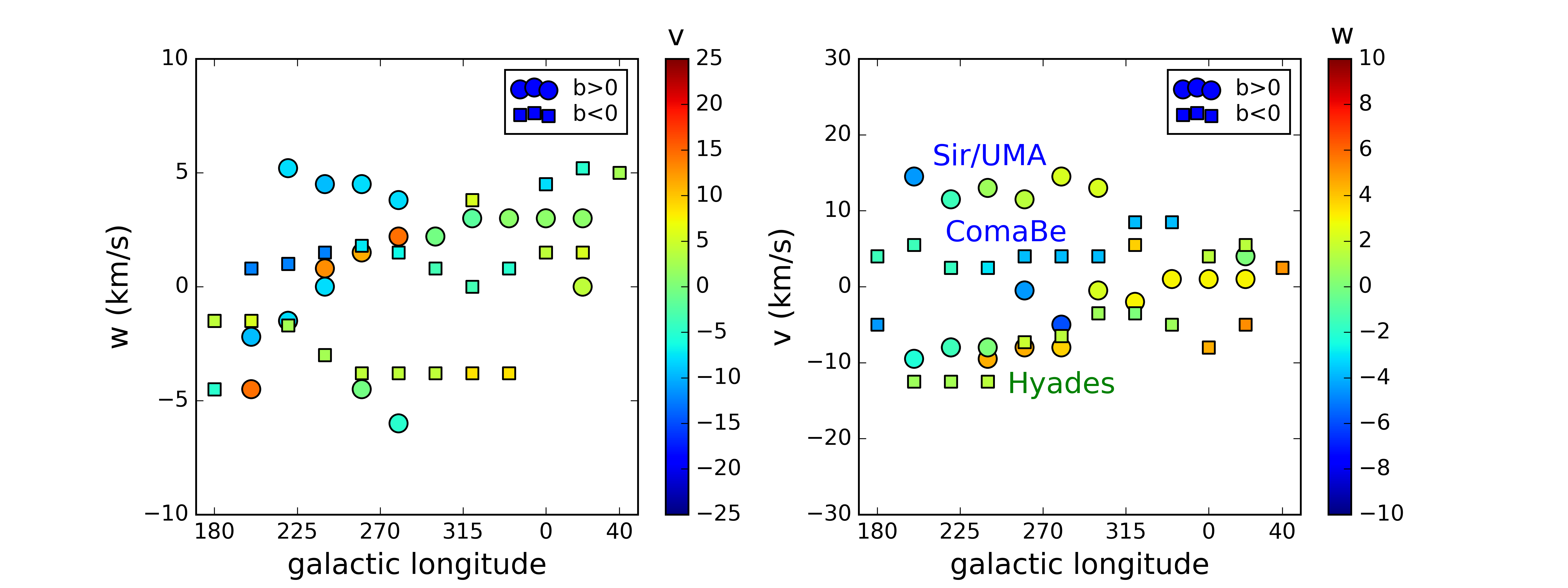}
\caption{Peaks seen in the $d>250$ pc ($v,w$) velocity distributions of near solar metallicity stars in
Figures \ref{fig:mmw_hem}
have $w = v_z$ velocity component plotted
as a function of galactic longitude in the left panel.  
The colour of the points depends on the $v$ 
velocity (with $v = -v_\theta- 230$ \kms, as before).
In the right panel, the peak $v$ velocity is plotted as a function of galactic longitude and with colour
set by the $w$ velocity. 
Sirius/UMa, Coma Berenices, and Hyades, 
streams or moving groups are most easily identified in the panel on the right where
they are labelled.  
This Figure is similar to Figure \ref{fig:uv_lon}
expect round points show the northern Galactic hemisphere (galactic latitude $b>0$) 
and square points show the southern Galactic hemisphere.
The Coma Berenices group is seen in the southern Galactic hemisphere whereas
the Sirius/UMa group is seen in the northern Galactic hemisphere.
The peak $w$ velocities of each group varies as a function of galactic longitude
and Galactic hemisphere.  Smooth variations in $w$ with longitude
are most easily seen in the right panel.
\label{fig:vw_lon}}
\end{figure*}

Vertical gradients in the ($u,v$) velocity distributions, if they are real, probably would be
associated 
with vertical velocity variations. To test this possibility, we examine velocity distributions
in ($v,w$) with $w = v_z$ the vertical velocity component.
Figure \ref{fig:mmw_hem} shows ($v,w$) velocity distributions for nearby ($d<250$ pc), near solar 
metallicity  stars (-0,1$<$[Fe/H]$<$0.2) that are in the north Galactic hemisphere
and that are in the southern hemisphere, separately.
Histograms are shown with $w$ on the x-axis and $v$ on the y-axis (top two panels) 
and vice versa for in bottom two panels.  The first and third panels show northern Galactic
hemisphere and the second and forth panels (from top) show the southern Galactic hemisphere.
Figure \ref{fig:mmw_hem}  shows that the $w$ values of velocity peaks  are 
dependent on Galactic longitude and on 
hemisphere.  For example, we compare the $\ell = 260^\circ$ histograms.
For $b>0$ the largest clump (associated with the Hyades group)  has $w \sim 3 $\kmsns,
whereas for $b<0$ the largest clump (with the Coma Berenices group)
has $w \sim -3 $ \kmsns.  These deviations are near but above our estimated precision 
of about 1 \kms.  The variations in peak velocities are shown as a function of Galactic longitude
in Figure \ref{fig:vw_lon}.  This figure is similar to Figure \ref{fig:uv_lon} except circles
show peaks from the velocity distributions in the northern Galactic hemisphere
(and identified in the velocity distributions of Figure \ref{fig:mmw_hem} top panel)
and squares show the northern Galactic hemisphere
(and identified in the velocity distributions of Figure \ref{fig:mmw_hem} bottom panel).

A single neighbourhood can contain more than one velocity
peak if  there are correlations in both vertical epicyclic and radial epicyclic phases.
Stars coming into the neighbourhood from the outer Galaxy could be at different
phases in the vertical oscillations  than stars moving outward from the inner Galaxy 
(see \citealt{delavega15}).
If the thin disc has a scale height of 300 pc and a vertical dispersion of 20 \kms 
(taking typical values from \citealt{bland16}),
a vertical velocity of $w = 3$ \kms would allow a star to 
reach about 45 pc above the Galactic plane.  This is high enough that the star might be seen
in one hemisphere and not the other.  In other words, the sizes of the vertical velocity
variations are approximately consistent with the extent
 of vertical motions that would allow differences in the velocity
distributions to be seen above and below the Galactic plane.
 
In summary, we find that the Coma Berenices group is predominantly seen at galactic
longitude $l \sim 230^\circ$, 
at galactic latitude $b<0$, and in solar and low metallicity stars.    
The Hyades and Sirius/UMa groups are more prominent at positive galactic latitude
in the same metallicity range and longitudes. 
The variations in velocity peak velocities in the velocity distributions
with Galactic hemisphere are not clearly associated
with variations in the stellar distance distributions but may be correlated with $\sim 3$ \kms variations
in peak vertical velocities $w$.  Such correlations are predicted if the Galactic disc hosts
vertical breathing and bending waves \citep{widrow12,widrow14,widrow15} or exhibits
correlated phases in vertical and radial epicyclic motions caused by perturbations in the outer Galaxy \citep{delavega15}.
However, asymmetries (with hemisphere) in both number density and velocity distribution
are largest where there are fewest
GALAH stars so the differences in the velocity distributions
with direction on the sky may be due to statistical variations.
Larger samples of stars are needed to see if the correlated variations in peak velocities in local
velocity distributions are real.

\section{Summary and Conclusion}

We have used
GALAH survey data  to construct local velocity distributions at neighbourhoods
near the Sun but comprised of stars seen at different galactic longitudes.
We find that the Hercules stream is most prominent in nearby high metallicity stars, with [Fe/H]$>0.2$,
(confirming the works by \citealt{liu15,liu16,perez17}), but is also
present in solar metallicity stars ($-0.1<$[Fe/H]$<0.2$), and difficult to see
in metal poor stars with  ([Fe/H]$<-0.1$).  An extension to low $v$ overlaying the Hercules stream
is also seen in high $\alpha$-element stars.
The dependence of the Hercules stream on metallicity is not surprising as metallicity
distributions at smaller galactocentric radii
contain more metal rich stars (e.g., \citealt{hayden15}).
  
We find that 
a gap in the velocity distributions
between the Hercules stream and stars in more nearly circular orbits
is approximately at an angular momentum value of $L_{gap} \approx 1640 \pm 40 $ \kmskpcns.
The gap is seen in a histogram of angular momentum versus orbital eccentricity
and is consistent with $v$ gap values in the local velocity distributions for neighbourhoods
of stars observed at different galactic longitudes.
The association of the Hercules stream with a particular angular momentum value
supports a bar resonant model as the angular momentum sets the period of orbits
and so determines those that are
resonant with the bar pattern speed.
The value we estimate for  $L_{gap}$ is consistent with that of  
the Outer Lindblad Resonance for a fast and short bar model  \citep{dehnen00,minchev07,antoja14,monari17,hunt18}, but 300 \kmskpc above that
of corotation for the slow and long bar model (that by \citealt{portail17a,perez17}).

When we construct velocity distributions for nearer stars with distance $d<250$ pc,
we recover peaks in the velocity distributions at locations of moving groups seen
previously in {\it Hipparcos} observations.   
However the location of these peaks
varies as a function of metallicity,
viewed galactic longitude and whether viewed in the northern or southern
Galactic hemispheres (positive or negative galactic latitudes).  
We infer that structures in the velocity distribution
varies over distances as short as a few hundred pc in the Galactic disc.
Gradients in peak velocities have size of order 25 \kmsikpcns, 
corresponding to changes of 10 \kms across 400 pc.
The variations over distances this short is unexpected.
Possible causes are streaming associated with multiple spiral density waves,
dissolved clusters,  resonances with bar or spiral patterns, 
or small scale spiral features such as feathers, branches, arm-segments or spurs.
We see differences in the $uv$ positions of velocity peaks between the two Galactic
hemispheres that may be correlated with $\sim 3$ \kms variations in
vertical velocity component $w$.   
Possible causes are 
vertical breathing and bending waves \citep{widrow12,widrow14,widrow15} or 
correlated phases in epicyclic motions caused by perturbations in the outer Galaxy \citep{delavega15}.
Unfortunately we see these variations at galactic longitudes where we have the fewest GALAH
stars and so suspect that small number statistics could have given us spurious peaks
in our normalized histograms. 

The velocity distributions we constructed for more distant stars (within 500 pc)
failed to show as much sub-structure as the nearby stars (within 250 pc). 
This could be due to coarse graining caused by using a larger volume to count 
stars or due to larger errors in the distances and space motions that are present
in the more distant stars.   The velocity distributions may depend on height from
the Galactic plane and may be biased by differences in the numbers of stars 
present in the survey along the different sight lines.
The GALAH survey stars are predominantly less than 1 kpc away.
%away and the survey contains fewer giants than other surveys such as RAVE.
Future studies with higher precision at greater distances will determine if substructure
in local velocity distributions exists in other regions in the Galaxy and if it varies
over short distances, as found here.   By studying in more detail
the abundances (chemical tagging; \citealt{desilva07,bland10,mitschang14,ting16}) 
using age indicators 
(such as chromospherically sensitive lines; \citealt{zerjal17}) and studying
larger and more uniformly distributed samples, we may differentiate
between causative dynamical mechanisms.

GALAH targets have a well defined magnitude distribution. 
A problem with a sample chosen with a narrow magnitude range
is that more distant stars tend to have higher luminosities.  
We have neglected this distance dependent difference in the stellar
distributions as we have no reason to think this would significantly impact the 
velocity or metallicity distributions.  We have not attempted to take into account 
how extinction affects the distributions of stars in the sample. 
Future attempts to characterize and compare local velocity distributions outside
the solar neighbourhood, at higher precision and at larger
distances from the Sun, may need to consider how variations in the sampled local stellar distributions
affect the local velocity and metallicity distributions.

\vskip 2 truein  % acknowledgements
A.C. Quillen thanks Mt. Stromlo Observatory
for their warm welcome and hospitality Nov 2017-- Feb 2018.
A.C. Quillen is grateful to the Leibniz Institut f\"ur Astrophysik Potsdam for their
warm welcome, support and hospitality July 2017 and May 2018.
A.C. Quillen is pleased to be a Visiting Scholar associated with the Hunstead Gift for Astrophysics
to the University of Sydney.
A. C. Quillen is grateful for generous support from the Simons Foundation and her work
 is in part supported by NASA grant 80NSSC17K0771.
We thank Agris Kalnajs for helpful and critical discussions.  Without his suggestion
we would not have found the results presented in Section \ref{subsec:hem}.

Parts of this research were conducted by the Australian Research Council (ARC) 
Centre of Excellence for All Sky Astrophysics in 3 Dimensions (ASTRO 3D), through
project number CE170100013.
Joss Bland-Hawthorn acknowledges a Miller Professorship 
from the Miller Institute, UC Berkeley,  and an ARC Laureate
Fellowship which also supports Gayandhi De Silva and Sanjib Sharma.
Sarah Martell acknowledges support from the ARC through DECRA Fellowship DE140100598.
Janez Kos is supported by an ARC DP grant awarded to Joss Bland-Hawthorn and Tim Bedding.  
Michael Hayden is supported by ASTRO-3D Centre 
of Excellence funding to the University of Sydney and an ARC DP grant awarded to Ken Freeman.
Ly Duong gratefully acknowledges a scholarship from Zonta International District 24. 
Ly Duong and Ken Freeman acknowledge support from ARC grant DP160103747.
Luca Casagrande is the recipient of an ARC Future Fellowship (project number FT160100402).

We acknowledge support of the Slovenian Research Agency (research core funding No. P1-0188).
David M. Nataf was supported by the Allan C. and Dorothy H. Davis Fellowship.
Yuan-Sen Ting is supported by the Carnegie-Princeton Fellowship and
the Martin A. and Helen Chooljian Membership from the Institute for Advanced Study in Princeton.

This work is
based on data acquired through the Australian Astronomical Observatory, under programmes: A/2013B/13 (The GALAH pilot survey); A/2014A/25, A/2015A/19, A2017A/18 (The GALAH survey); A/2015A/03, A/2015B/19, A/2016A/22, A/2016B/12, A/2017A/14 (The K2-HERMES K2-follow-up program); A/2016B/10 (The HERMES-TESS program); A/2015B/01 (Accurate physical parameters of Kepler K2 planet search targets); S/2015A/012 (Planets in clusters with K2).

\vskip 0.5 truein1

\appendix

\section{Epicyclic Motion}
\label{sec:dyn}

Taking into account motions in the galactic plane only, the energy per unit mass of a star
\begin{equation}
E = \frac{v_r^2}{2} + \frac{v_\theta^2}{2} + v_c^2 \ln \frac{r}{R_\odot}, \label{eqn:EE}
\end{equation}
where we have assumed a flat rotation curve with velocity $v_c$ and 
neglected bar and 
spiral arm perturbations in the gravitational potential.
A constant offset to the energy is chosen so a circular orbit with radius $R_\odot$ has energy 
$E={v_c^2}/{2}$.
%We now work with velocity in units of circular velocity $v_c$ and radius in units of $R_\odot$.
The $z$ component of angular momentum per unit mass of a star is 
\begin{equation}
L = r v_\theta,
\end{equation}
where $r$ is the star's galactocentric radius and $v_\theta$ the tangential velocity component.

Following \citet{quillen14}, (also see \citealt{fuchs01}  and \citealt{cont75}), 
a Hamiltonian describing planar motion in an axisymmetric galaxy in action angle variables is
\begin{equation}
H(L,\theta, J, \varphi) = E_c(L) + \kappa(L) J     \label{eqn:H0}
\end{equation}
to first order in epicyclic action variable $J$.
The function $E_c(L)$ was used by \citet{dehnen99} in the development of
an accurate low order epicyclic approximation.  
Here the angular momentum $L$ is canonically conjugate to the azimuthal angle $\theta$
and the radial action variable $J$ is conjugate to the epicyclic angle $\varphi$.
The function $E_c(L)$ gives the angular rotation rate $\Omega(L)$ for a circular orbit via
Hamilton's equation, or 
\begin{equation}
\dot \theta = \Omega(L) = \frac{\partial H}{\partial L} (J=0) = \frac{\partial E_c(L)}{\partial L}.
\end{equation}
The frequency $\kappa(L)$ is the epicyclic frequency of
a nearly circular orbit with angular momentum $L$.

A circular orbit with angular momentum $L$ has radius  $r_c(L)$.
Epicyclic oscillations are related to the radial action angle variables
\begin{align}
r &= r_c(L) + \sqrt{\frac{2J}{\kappa(L)}} \cos \varphi  \label{eqn:rorb} \\
\dot r &= % p_r = 
- \sqrt{2 J \kappa(L)} \sin \varphi \label{eqn:rJ} \\
\dot \varphi &= \kappa(L).
\end{align}
This convention for epicyclic angle gives apocentre at $\varphi=0$, pericentre at $\varphi=\pi$, 
radial velocity $v_r<0$ at $\varphi=\pi/2$
and $v_r>0$ at $\varpi = -\pi/2$.  The Hercules stream with positive $v_r$ would have
epicyclic angle $-\pi/2 < \varphi < 0$.  The Hercules stream has stars that have just passed their pericentre
and are moving toward apocentre.  
From equation \ref{eqn:rorb} we can define an orbital eccentricity
\begin{equation}
e \equiv \frac{1}{r_c(L)} \sqrt{ \frac{2 J}{\kappa(L)}}. \label{eqn:ecc}
\end{equation}

For a flat rotation curve with circular velocity $v_c$
\begin{eqnarray}
r_c(L) &=& \frac{L}{v_c}  \label{eqn:rL} \\
\Omega(L) &=&  \frac{v_c^2}{L} \\
\kappa(L) &=& \sqrt{2} \frac{v_c^2}{L}  \label{eqn:kappaL} \\ 
E_c(L) &=& v_c^2 \left( \frac{1}{2} + \ln \frac{L}{L_\odot} \right), \label{eqn:EL}
\end{eqnarray}
with $L_\odot = R_\odot v_c$.

The difference between energy (equation \ref{eqn:EE}) and that of a circular orbit
(equation \ref{eqn:EL}) is related to the orbital eccentricity
\begin{equation}
\frac{E - E_c(L)}{v_c^2} = \frac{\kappa(L) J}{v_c^2} =  e^2 \label{eqn:energy_ee}
\end{equation}
where we have used equations \ref{eqn:H0}, \ref{eqn:ecc}, \ref{eqn:kappaL} and \ref{eqn:EL}.

\section{Distance Distributions}
\label{sec:dist}

\begin{figure*}
\includegraphics[width=10.5cm]{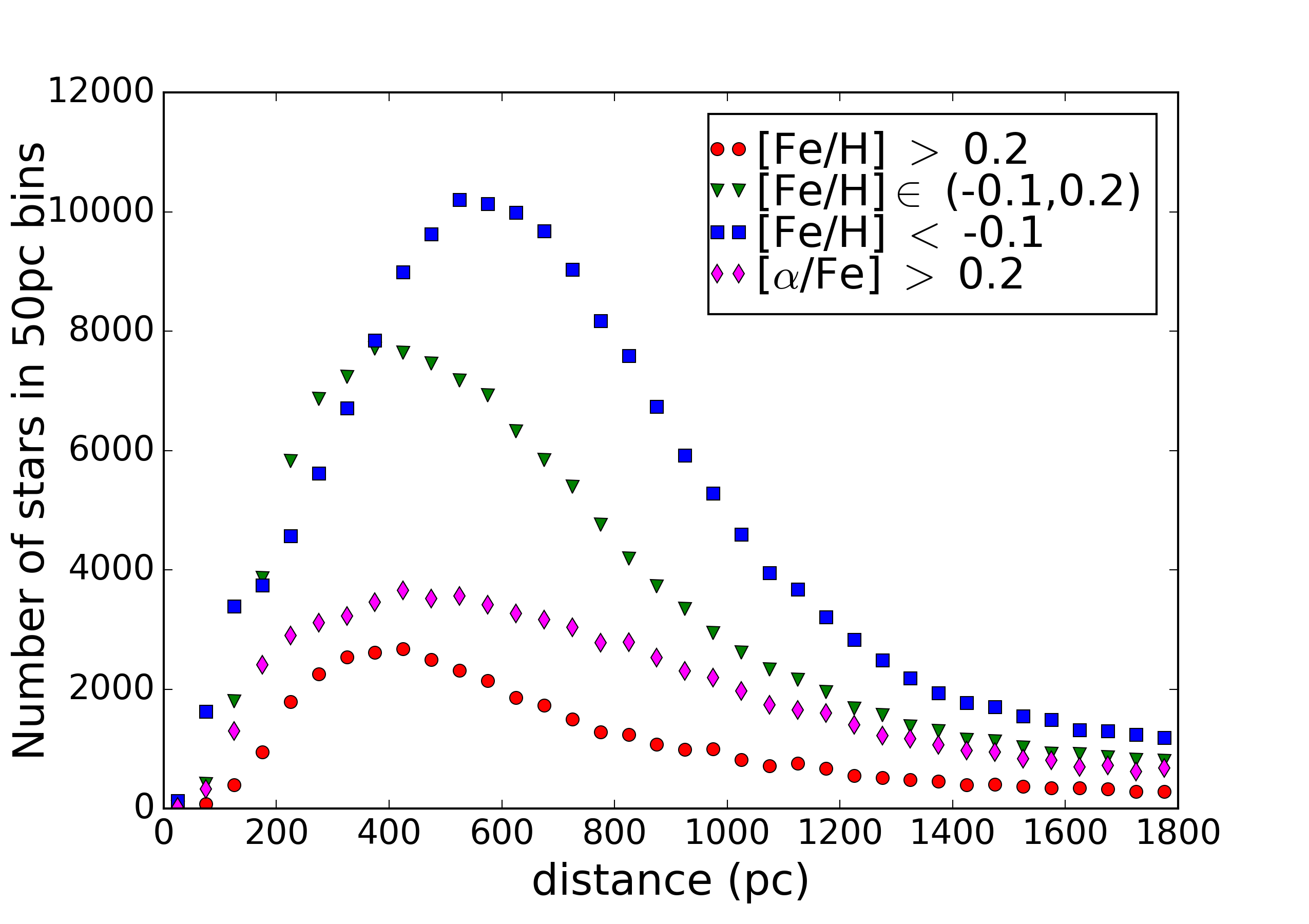}
\caption{The distribution of distances for the 4 sets of stars used
to construct velocity distributions shown in Figures \ref{fig:line}-- \ref{fig:nn_line}.
\label{fig:dist_hist}}
\end{figure*}

In Figure \ref{fig:dist_hist}
we show the distribution of distances for the different samples of GALAH stars
used to make the histograms shown in Figures \ref{fig:line}--\ref{fig:nn_line}. 
Stars at all galactic longitudes are included in these histograms.   The distribution
shows the numbers of stars in 50 pc bins for high metallicity [Fe/H] $>0.2$ (red circles),
near solar metallicity
$-0.1 <$ [Fe/H] $< 0.2$ (green triangles), 
lower metallicity [Fe/H] $<-0.1$ (blue squares), and high $\alpha$-element
[$\alpha$/Fe]$ > 0.2$ and $T_{eff}>4500^\circ$ (magenta diamonds).  The distance distributions are
consistent with the GALAH sample being dominated by dwarf stars rather than giants.

\FloatBarrier

\section{Cuts to GALAH survey stars}
\label{sec:ap_cuts}

We list in Table \ref{tab:cuts} the restrictions and cuts made 
to the GALAH survey stars when creating figures and Tables.

\begin{table*}
\vbox to 85mm{\vfil
\caption{\large  List of cuts in the survey data used to make Figures and Tables
\label{tab:cuts}}
\begin{tabular}{@{}llllll}
Description & Restriction & Figures  & Tables \\
\hline
High galactic latitude cut & $|b| < 45^\circ$ & All &  \ref{tab:tab1}--\ref{tab:tab4} \\
Fields$^a$   & $0< $ field\_id $ \le 7338$,  & All & 
         \ref{tab:tab1}--\ref{tab:tab4} \\
     &  \ \  or field\_id $\ge 7357$ & & \\
Distance errors  & $\sigma_d/d < 0.5$ & All &   \ref{tab:tab1}--\ref{tab:tab4} \\
Supersolar metallicities & [Fe/H] $> 0.2$ &  \ref{fig:line}a, \ref{fig:vhist}, 
	\ref{fig:LE} top panel, \ref{fig:fn_line}, \ref{fig:vhist_line},
    \ref{fig:nn_line}a, \ref{fig:nn_peak}a, \ref{fig:arrow}--\ref{fig:fnn_b}, \ref{fig:dgl}a,  
      \ref{fig:dist_hist} &  \ref{tab:tab1} \\
Near solar metallicities & $-0.1 < $[Fe/H] $<0.2$ &  \ref{fig:line}b, 
	\ref{fig:vhist}, \ref{fig:nn_line}b, \ref{fig:nn_peak}b, 
       \ref{fig:arrow}, \ref{fig:uv_lon}, \ref{fig:mnn_b}, \ref{fig:dgl}b,
      \ref{fig:mmw_hem}, \ref{fig:vw_lon}, \ref{fig:dist_hist} &   \ref{tab:tab2}, \ref{tab:tab4}  \\
Subsolar metallicities &   [Fe/H] $< -0.1$ &  \ref{fig:line}c, \ref{fig:vhist},
	 \ref{fig:LE} midpanel, \ref{fig:nn_line}c, 	
	\ref{fig:nn_peak}c,   \ref{fig:arrow}, \ref{fig:uv_lon}, 
	\ref{fig:lnn_b}, \ref{fig:dgl}c, \ref{fig:dist_hist} & \ref{tab:tab3} \\
$\alpha$-rich      &  [$\alpha$/Fe] $> 0.2$  & \ref{fig:line}d, \ref{fig:vhist},
	 \ref{fig:LE}  bottom panel, \ref{fig:dist_hist} & none \\
Effective temperature     & $T_{\rm eff} > 4500^\circ$ K & \ref{fig:line}d, \ref{fig:vhist}, 
	\ref{fig:LE}  bottom panel, \ref{fig:dist_hist} & none \\
Distance  & $0<d<500$ pc &  \ref{fig:line}, \ref{fig:fn_line}, \ref{fig:vhist_line}  & none\\
Distance  & $0<d<1000$ pc & \ref{fig:vhist}, \ref{fig:LE} & none \\
Distance  & $0<d<250$ pc &  \ref{fig:nn_line}--\ref{fig:vw_lon} &   \ref{tab:tab1}--\ref{tab:tab4} \\
Northern Galactic Hemisphere  & $0 < b < 45^\circ$ &  
       \ref{fig:fnn_b}a, \ref{fig:mnn_b}a, \ref{fig:lnn_b}a, \ref{fig:dgl} top panels,
        \ref{fig:mmw_hem}a,c, \ref{fig:vw_lon} &  \ref{tab:tab4} top \\
Southern Galactic  Hemisphere & $0 > b > -45^\circ$ &  
       \ref{fig:fnn_b}b, \ref{fig:mnn_b}b, \ref{fig:lnn_b}b, \ref{fig:dgl} middle panels,
        \ref{fig:mmw_hem}b,d, \ref{fig:vw_lon} & \ref{tab:tab4} bottom \\
M67 field removed$^b$  & field\_id $\ne 6605$ & 
	\ref{fig:nn_line}b, \ref{fig:nn_peak}b, 
       \ref{fig:arrow}, \ref{fig:uv_lon}, \ref{fig:mnn_b}, \ref{fig:dgl}b,
      \ref{fig:mmw_hem}, \ref{fig:vw_lon}
&  \ref{tab:tab2}, \ref{tab:tab4} \\
Radial velocity component & $-40 < -v_r < 10$ \kms & \ref{fig:vhist}, \ref{fig:vhist_line} & none \\
%Radial velocity component & $-40 < -v_r < 10$ \kms & \ref{fig:vhist_line} & none \\
Galactic longitude & $250^\circ < \ell < 290^\circ$ & \ref{fig:vhist} & none \\
\hline
 \end{tabular}
{\\  Notes. 
$^a$Pilot survey and open cluster target fields are excluded. 
$^b$Stars in the field containing M67 observed as part of the K2-HERMES observing program are excluded
when making histograms of nearby ($d<250$ pc)  and near solar metallicity 
stars ($-0.1 < $[Fe/H] $<0.2$).
}}
\end{table*}

\FloatBarrier

\section{Velocities of Peaks in Histograms}
\label{sec:ap_peaks}

In Tables \ref{tab:tab1} -- \ref{tab:tab3} 
we show the ($u,v$) velocity components of peaks identified in the velocity distributions
of Figure \ref{fig:nn_line}
and that
that are shown in Figures \ref{fig:nn_peak}--\ref{fig:uv_lon}.
Table \ref{tab:tab1} is for higher metallicity stars, Table \ref{tab:tab2} is for near solar metallicity stars
and Table  \ref{tab:tab3} is for lower metallicity stars.
The conventions for computing ($u,v$) (i.e., solar peculiar velocity and LSR
circular velocity)  are described in Section \ref{sec:coord}.
In these tables,
galactic longitudes are in degrees and velocities are in \kmsns.  
We have tentatively labelled peaks by their proximity to previously identified moving groups or streams
(see Section \ref{subsec:peaks}). 
%We use ComaBe to denote velocities near that of the Coma Berenices group.
The peaks are identified in histograms that have 4 \kms square
bins in ($u,v$) and 
that were  smoothed with a Gaussian filter with standard deviation 4 \kms wide.   
The precision of these velocity measurements is about $\pm$ 3 \kmsns.
The peak height in these tables is in units of numbers of stars in the bin
divided by the bin area (16 (\kmsns)$^2$) and divided by the number of stars in the histogram.

Table \ref{tab:tab4} shows ($w,v$) velocity components of peaks identified in the velocity distributions
of Figures \ref{fig:mmw_hem} and shown in Figure \ref{fig:vw_lon} 
for near solar stars metallicity, with $d>250$ pc, and for stars in the two  Galactic
hemispheres.
Table \ref{tab:tab4} is similar to Tables \ref{tab:tab1} -- \ref{tab:tab3}
except the bins were 3 \kms wide in $v$ and 1.5 \kms wide in $w$.
The peak height in this table is in units of numbers of stars in the bin
divided by the bin area (4.5 (\kmsns)$^2$) and divided by the number of stars in the histogram.
The top part of the table shows the northern Galactic hemisphere (latitude $b>0$)
and the bottom half shows the southern Galactic hemisphere.

\begin{table}
\vbox to140mm{\vfil
\caption{\large  Peaks in  [Fe/H]$>0.2$ and $d<250$ pc ($u,v$) Histograms \label{tab:tab1}}
\begin{tabular}{@{}rrrllll}
\hline
%[Fe/H]>0.2 
  $\ell$     &   $u$   &    $v$   &  height      & stream/group \\
  ($^\circ$) & (\kmsns)   & (\kmsns)     &   & \\
\hline
   180    &   0   &    0   & 0.3  & \\
   180    & -26   &   -4   & 0.3  & Hyades  \\
   180    &   8   &  -12   & 0.2  &  \\
   200    & -28   &   -8   & 0.8  & Hyades \\
   220    & -28   &   -8   & 0.7  & Hyades \\
   220    &  14   &   -8   & 0.3  & Pleiades \\
   240    & -10   &  -10   & 0.4  & Hyades \\
   240    &  24   &  -12   & 0.2  & Pleiades \\
   260    & -12   &   -6   & 0.8  & Hyades \\
   260    &  24   &  -10   & 0.4  & Wolf630 \\
   260    & -12   &  -37   & 0.4  & Hercules \\
   260    &  14   &  -43   & 0.2  & HR1614 \\
   280    & -10   &   -6   & 1.3 & Hyades \\
   280    &  -4   &  -33   & 0.7 & Hercules \\
   280    &  18   &  -41   & 0.4 & HR1614 \\
   300    &  26   &   -6   & 0.6  & Wolf630 \\
   300    & -10   &   -4   & 1.1  & Hyades \\
   300    &  -4   &  -31   & 0.6  & Hercules \\
   300    &  20   &  -39   & 0.3  & HR1614 \\
   320    &  30   &   -6   & 0.5  & Wolf630 \\
   320    & -20   &   -2   & 0.7  & Hyades \\
   320    &  10   &    4   & 0.5  & \\
   320    &  -6   &  -31   & 0.4  & Hercules \\
   340    & -22   &   -2   & 0.8 & Hyades \\
   340    &  34   &   -6   & 0.5 & Wolf630 \\
   340    &  -2   &  -33   & 0.3 & Hercules \\
     0    & -18   &   -0   & 0.6 & \\
     0    &   2   &    4   & 0.5 & \\
    20    & -18   &   -0   & 0.7 & \\
    20    &   2   &    4   & 0.5 & \\
    \hline
 \end{tabular}
{\\ 
}}
\end{table}

\begin{table}
\vbox to170mm{\vfil
\caption{\large  Peaks in  $-0.1<$[Fe/H]$<0.2$ and $d<250$ pc ($u,v$) Histograms \label{tab:tab2}}
\begin{tabular}{@{}rrrllll}
\hline
  $\ell$     &   $u$   &    $v$   &  height      & stream/group \\
    ($^\circ$) & (\kmsns)   & (\kmsns)     &   & \\
\hline
%
%-0.1<[Fe/H]<0.2
%   l     &   u   &    v   &  N      & stream/group \\
   180    &  -4   &    6   & 1.2 & \\
   180    &  17   &    2   & 0.7 & \\
   180    & -22   &   -5   & 0.9 & Hyades \\
   180    & -10   &   -8   & 1.0 & Hyades/Pleiades \\
   200    &  20   &   11   & 0.9 & Sirius/UMA \\
   200    &  -3   &    4   & 1.3 & Coma Berenices \\
   200    &  13   &   -1   & 0.8 & \\
   200    & -29   &   -8   & 2.4 & Hyades \\
   220    &  19   &   11   & 1.0 & Sirius/UMA \\
   220    &   1    &    4   & 1.3 & Coma Berenices \\
   220    & -15   &   -8   & 1.4 & Pleiades \\
   220    & -29   &   -8   & 2.4 & Hyades \\
   240    &  18   &   13   & 1.2 & Sirius/UMA \\
   240    &   1   &    2   & 1.4 & Coma Berenices \\
   240    &  -8   &   -8   & 1.3 & Hyades \\
   260    &  20   &   13   & 2.2 & Sirius/UMA \\
   260    &   1   &    4   & 2.9 & Coma Berenices \\
   260    & -24   &   -5   & 1.7 & Hyades \\
   260    &  -6   &   -7   & 2.6 & Pleiades \\
   260    & -10   &  -33   & 0.8 & Hercules \\
   280    &  20   &   13   & 2.7 & Sirius/UMA \\
   280    &   3   &    4   & 3.3 & Coma Berenices \\
   280    & -10   &   -3   & 3.4 & Hyades \\
   280    &   1   &   -8   & 3.2 & Pleiades \\
   280    &  -8   &  -30   & 1.2 & Hercules \\
   300    &  20   &   13   & 2.4 & Sirius/UMA \\
   300    & -10   &   -1   & 3.4 & Hyades \\
   300    &   3   &   -7   & 2.9 & Pleiades \\
   300    & -10   &  -30   & 1.0 & Hercules \\
   320    &  -6   &   -1   & 2.6 & Hyades/Pleaides \\
   320    &  17   &   14   & 1.8 & Sirius/UMA \\
   340    & -18   &    0   & 2.3  & \\
   340    &   4   &    4   & 2.6  & \\
   340    &  24   &   -3   & 1.3  & \\
   340    & -17   &  -21   & 1.0  & \\
     0    &   4   &    2   & 1.1  & \\
     0    & -20   &    4   & 1.0  & \\
     0    &  18   &    6   & 0.7  & \\
     0    & -10   &    7   & 1.0  & \\
     0    &  11   &  -14   & 0.6  & \\
    20    & -19   &    7   & 1.2  & \\
    20    &   4   &    2   & 1.3  & \\
    20    & -18   &    4   & 1.2  & \\
    20    &  -6   &   -5   & 1.2  & \\
    40    &   6   &    6   & 0.6  & \\
    40    &  -8   &   -3   & 0.5  & \\
    \hline
 \end{tabular}
{\\ 
}}
\end{table}

\begin{table}
\vbox to100mm{\vfil
\caption{\large  Peaks in  [Fe/H]$<-0.2$ and $d<250$ pc ($u,v$) Histograms \label{tab:tab3}}
\begin{tabular}{@{}rrrllll}
\hline
  $\ell$     &   $u$   &    $v$   &  height      & stream/group \\
    ($^\circ$) & (\kmsns)   & (\kmsns)     &   & \\
\hline
%[Fe/H]<-0.1
%   l     &   u   &    v   &  N      & stream/group \\
   180    &  -2   &    2   & 1.4 & \\
   180    & -20   &   -4   & 1.1 & Hyades \\
   200    &  14   &   10   & 1.2 & Sirius/UMA \\
   200    &  -4   &    2   & 1.4 & Coma Berenices \\
   200    & -22   &   -8   & 1.4 & Hyades \\
   220    &  18   &   10   & 1.1 & Sirius/UMA \\
   220    &  -6   &    2   & 1.1 & Coma Berenices \\
   220    & -20   &  -10   & 1.4 & Hyades \\
   240    &  14   &   12   & 1.0 & Sirius/UMA \\
   240    &  -2   &    2   & 0.9 & Coma Berenices \\
   240    & -16   &   -8   & 1.0 & Hyades \\
   240    &   4   &   -8   & 1.0 & Pleiades \\
   260    &  12   &   14   & 2.3 & Sirius/UMA \\
   260    &   0   &    0   & 2.3 & Coma Berenices \\
   260    &   4   &  -12   & 2.3 & Pleiades \\
   280    &   2   &    2   & 3.4 & \\
   300    &   6   &   -4   & 3.5 & \\
   320    &   8   &   -4   & 3.0 & \\
   340    &   6   &    0   & 2.7 & \\
   340    & -10   &    6   & 2.3 & \\
     0    &  -2   &    6   & 1.2 & \\
    20    &   0   &    6   & 1.3 & \\
    40    &  18   &   18   & 0.5 & \\
    \hline
 \end{tabular}
{\\ 
}}
\end{table}

\begin{table}
\vbox to200mm{\vfil
\caption{\large  Peaks in  $-0.1<$[Fe/H]$<0.2$ and $d<250$ pc ($w,v$) Histograms 
with Galactic Hemisphere \label{tab:tab4}}
\begin{tabular}{@{}rrrllll}
\hline
  $\ell$     &   $w$   &    $v$   &  height      & stream/group \\
    ($^\circ$) & (\kmsns)   & (\kmsns)     &   & \\
\hline
\multicolumn{2}{l}{For Galactic latitude $b>0$} &&& \\
\hline
%             w        v 
   200  &    -4.5  &  14.5 &  0.4 & Sirius/UMa \\
   200  &    -2.2  &  -9.5 &  1.2 & Hyades/Pleiades \\
   220  &    -1.5  &  11.5 &  0.5 & Sirius/UMa  \\
   220  &     5.2  &  -8.0 &  1.1 &\\
   220  &    -1.5  &  -8.0 &  1.3 & Hyades/Pleiades \\
   240  &     0.8  &  13   &  0.5 &  Sirius/UMa  \\
   240  &     0.0  &  -8.0 &  0.5 &\\
   240  &     4.5  &  -9.5 &  0.6 & Hyades/Pleiades  \\
   260  &     1.5  &  11.5 &  0.6 &  Sirius/UMa   \\
   260  &    -4.5  &  -0.5 &  0.6 &\\
   260  &     4.5  &  -8.0 &  1.1 & Hyades/Pleiades  \\
   280  &     2.2  &  14.5 &  0.6 &  Sirius/UMa  \\
   280  &    -6.0  &  -5.0 &  0.8 &\\
   280  &     3.8  &  -8.0 &  1.3 & Hyades/Pleiades  \\
   300  &     2.2  &  13.0 &  0.8 & Sirius/UMa  \\
   300  &     2.2  &  -0.5 &  1.5 &  \\
   320  &     3.0  &  -2.0 &  1.6 &\\
   340  &     3.0  &   1.0 &  1.7 &\\
     0  &     3.0  &   1.0 &  0.6 &\\
    20  &     0.0  &   4.0 &  0.7 &\\
    20  &     3.0  &   1.0 &  0.7 &\\
%
% b<0         w        v 
\hline
\multicolumn{2}{l}{For Galactic latitude $b<0$} &&& \\
\hline
%  l         w         v    N
   180   &   -1.5 &     4.0 & 1.2 &  Coma Berenices\\
   180   &   -4.5  &   -5.0 & 1.0 &  \\
   200   &   -1.5  &    5.5 & 0.8 & Coma Berenices \\
   200   &    0.8  &  -12.5 & 0.6 & Hyades/Pleiades \\
   220   &   -1.7  &    2.5 & 0.6 & Coma Berenices \\
   220   &     1.0 &   -12.5 & 0.6 & Hyades/Pleiades \\
   240   &   -3.0  &    2.5 & 0.8 &Coma Berenices \\
   240   &    1.5   & -12.5 & 0.6 & Hyades/Pleiades \\
   260   &    -3.8  &    4.0 & 1.6 & Coma Berenices \\
   260   &     1.8  &   -7.3 & 1.3 & Hyades/Pleiades  \\
   280   &   -3.8   &   4.0 & 1.6& Coma Berenices \\
   280    &   1.5   &  -6.5 & 2.0&  \\
   300    &   -3.8  &   4.0 & 1.3& \\
   300    &    0.8  &  -3.5 & 1.5& \\
   320    &   -3.8   &   8.5 & 0.9  \\
   320     &   3.8   &   5.5 & 1.0 \\
   320    &    0.0   &  -3.5 & 1.1 \\
   340   &    -3.8   &   8.5 & 0.9 \\
   340    &    0.8  &   -5.0 & 0.9 \\
     0     &   1.5    &  4.0 & 0.7 \\
     0    &    4.5   &  -8.0 & 0.6 \\
    20   &     1.5  &    5.5 & 0.7 \\
    20   &     5.2  &   -5.0 & 0.6 \\
    40    &    5.0    &    2.5 & 0.4 \\
\hline
 \end{tabular}
{\\ 
}}
\end{table}

%\FloatBarrier

\end{document}